\documentclass[hidelinks,review,numbers,sort&compress,fleqn,final,3p,times]{elsarticle}

\usepackage[labelfont=bf,justification=raggedright,singlelinecheck=false]{caption}
\AtBeginDocument{%

\def\equationautorefname~#1\null{Eq.~(#1)\null}
}
\usepackage{natbib}
\setcitestyle{square,sort&compress,comma,numbers}
\usepackage{xcolor}
\usepackage{tabularx}
\usepackage{comment}
\usepackage[unicode]{hyperref}
    \hypersetup{
        colorlinks   = true,
        citecolor    = blue
    }
\usepackage{amssymb}
\usepackage{pifont}
\usepackage{mathtools, cuted}
\usepackage{subcaption}  
\usepackage{anyfontsize} 
\usepackage{threeparttable}
\usepackage{rotating}
\usepackage{multicol}
\usepackage{graphicx}
\usepackage{epsfig}
\usepackage{booktabs}

\usepackage{amsthm}
\usepackage{float}

\journal{Chinese Journal of Physics}

\begin{document}

\begin{frontmatter}


\title{Dark Energy Stars in Finch--Skea Spacetime with a Schwarzschild--(Anti--)de Sitter Exterior}
\author[]{Muhamad Ashraf Azman\corref{cor1}}
\ead{asaz@ukm.edu.my}
\cortext[cor1]{Corresponding author}
\affiliation{organization={Department of Applied Physics, Faculty of Science and Technology},
     addressline={Universiti Kebangsaan Malaysia},
    postcode={43600}, 
    state={Bangi, Selangor},
    country={Malaysia}}



\begin{abstract}
In this work, the effects of the cosmological constant on dark energy stars in Finch--Skea spacetime are systematically investigated within the framework of Einstein gravity. A static and anisotropic stellar configuration composed of ordinary matter and dark energy is constructed by employing the complexity factor formalism to obtain the temporal metric potential. The interior solution is smoothly matched to the exterior Schwarzschild--(anti--)de Sitter spacetime through appropriate boundary conditions. The influence of positive and negative cosmological constants on the structural and stability properties of the compact star candidate Vela X-1 is investigated in detail. The results show that positive values of the cosmological constant produce larger and less compact configurations, whereas negative values lead to denser and more compact stars with stronger gravitational effects. The energy conditions are satisfied throughout the stellar interior for all considered cases. However, sufficiently large values of the cosmological constant introduce deviations from hydrostatic equilibrium and affect the causal and dynamical stability behavior of the model. Overall, the results demonstrate that the cosmological constant plays an important role in determining the compactness, equilibrium and stability behavior of dark energy stars in Finch--Skea spacetime.
\end{abstract}



\begin{keyword}
Dark energy stars \sep Cosmological constant \sep Finch--Skea spacetime \sep Schwarzschild--de Sitter spacetime \sep Schwarzschild--anti--de Sitter spacetime \sep Compact stars



\end{keyword}

\end{frontmatter}



\section{Introduction}\label{sec:1}

The study of compact stars has remained an active and enduring area of research in theoretical astrophysics and general relativity since the pioneering work of Ref. \cite{PhysRev.55.374}, who first proposed a relativistic model of a neutron star based on a degenerate fermionic equation of state. Over the decades, this field has significantly evolved to encompass a wide variety of compact astrophysical objects, including white dwarfs, neutron stars and black holes. The motivation for studying such objects stems from their extreme physical conditions—high densities, strong gravitational fields and relativistic effects—that serve as natural laboratories for testing fundamental theories of matter and gravity. In response to increasingly precise observational data, particularly from radio pulsars, X-ray binaries and gravitational-wave detections, numerous theoretical models have been developed to understand the behavior of the external gravitational field of compact stars and to describe their internal structure. 

One of the interesting theoretical models of compact stars is based on the assumption that dark energy exists within the stellar configuration. Such objects are commonly referred to as dark energy stars and are commonly described by the equation of state \(p_r = \omega \rho\). This model was first proposed in Ref.~\cite{2005tsra.conf..101C} and has since been extensively investigated from various directions.  

One important direction in the study of dark energy stars concerns the choice of interior metric potentials. Various forms of metric potentials have been considered, including the Krori--Barua \cite{2024ChJPh..92.1703A,2012GReGr..44..107R,2015Ap&SS.356...57D,Bhar2023ExistenceGravity,2024PDU....4301398D}, Tolman-IV \cite{2021ChJPh..71..683D,AZMAN2025170242}, Tolman--Kuchowicz \cite{2021PDU....3400879B,RN865,RN866} and Finch--Skea \cite{2024ApSS.369...76D,2020MPLA...3550071B,2024ChJPh..87..608R,2022arXiv220613943M} metric potentials, where different assumptions regarding the interior spacetime geometry lead to a wide variety of dark energy star models. Nevertheless, several common features persist among these models, such as negative pressure profiles and the violation of the strong energy condition, both of which are often associated with instability. Consequently, some researchers have incorporated ordinary matter components into the equation of state through a coupling parameter in order to construct more stable stellar configurations.

Beyond Einstein gravity, such stellar configurations have also been investigated in modified gravitational frameworks, including Gravity's Rainbow \cite{RN691}, Einstein--Maxwell gravity \cite{2015Ap&SS.356...57D}, Rastall gravity \cite{RN865}, $f(\mathcal{Q})$ gravity \cite{Bhar2023ExistenceGravity,RN866} and $f(\mathcal{R},\mathcal{G})$ gravity \cite{2024PDU....4301398D}. These investigations provide further insight into the reproducibility of dark energy star models within different gravitational frameworks. In many cases, the resulting stellar properties remain largely indistinguishable across the various theories, demonstrating the viability and relevance of these alternative gravitational frameworks beyond the conventional Einsteinian description.

Recently, Ref.~\cite{2024ApSS.369...76D} introduced a new direction in the study of dark energy stars by incorporating the cosmological constant \(\Lambda\) (vacuum energy) into the model through the assumption of an exterior Schwarzschild--de Sitter spacetime. This approach is relatively novel, since most previous studies considered only the exterior Schwarzschild solution or assumed \(\Lambda = 0\). Such a framework may provide a possible mechanism for establishing a connection between the interior and exterior spacetime through the dark energy fluid distribution itself \cite{Farrah2023ObservationalEnergy}. 

Despite these developments, several aspects of the study remain unexplored. Although a nonzero cosmological constant has previously been incorporated into dark energy star models, its influence on the internal structure and physical behavior of the configuration has not been examined in detail. This issue becomes particularly relevant in light of recent observational and phenomenological studies suggesting possible deviations from a strictly constant cosmological term \cite{10.1093/mnras/staf1685,tr6y-kpc6,z2q4-qcdq}. Consequently, it is important to investigate how variations in $\Lambda$ influence the physical properties and stability behavior of dark energy stars within Finch--Skea spacetime.

In the present work, a Finch--Skea type metric potential similar to that considered in Ref.~\cite{2024ApSS.369...76D} is adopted, but with a different functional form. In particular, the $g_{rr}$ metric potential from Ref.~\cite{2020MPLA...3550071B} is employed, while the temporal metric potential $g_{tt}$ is obtained through the complexity factor formalism developed in Ref.~\cite{2024ChJPh..87..608R}. A summary of the applications of Finch--Skea metric potentials in compact star studies is presented in \autoref{tab:summary-finch-skea-metric}, illustrating the versatility of this spacetime geometry in modeling anisotropic stars \cite{2025ApSS.370...46I, 2023ChPhC..47a5104S,2023GrCo...29..206J,2020CQGra..37g5017D,2021JApA...42...74B,2023ChJPh..81..362D}, perfect fluid stars \cite{2013GReGr..45..717B,2021IJGMM..1850160B}, gravastars \cite{MAJEED2022101802,2023MPLA...3850123S}, quark stars \cite{2013IJTP...52.3319K} and dark energy stars.

The present study differs from previous Finch--Skea-based dark energy star models in several important aspects. First, Schwarzschild, Schwarzschild--de Sitter and Schwarzschild--anti--de Sitter exterior geometries are systematically compared within a unified framework. Second, the effects of both positive and negative cosmological constants on the physical viability and stability properties of the stellar configuration are investigated in detail. Third, the present model is systematically compared with previous dark energy star models constructed using Finch--Skea metric potentials.

The article is organized as follows. \autoref{sec:2} presents the interior spacetime geometry and the corresponding Einstein field equations governing the system composed of ordinary matter and dark energy. \autoref{sec:3} develops the stellar model by introducing the equation of state, the complexity factor formalism used to derive the metric potential $e^{\nu}$ and thermodynamic quantities. In \autoref{sec:4}, the interior spacetime is smoothly matched to the exterior Schwarzschild--(anti--)de Sitter geometry through the appropriate boundary conditions, leading to the determination of the unknown model parameters. \autoref{sec:5} provides a detailed investigation of the physical properties and stability of the resulting compact configuration, including the behavior of the metric potentials, thermodynamic matter profiles, anisotropy, mass function, compactness, redshift, energy conditions, hydrostatic equilibrium, causality, cracking condition and adiabatic stability. Finally, \autoref{sec:6} summarizes the main findings of the study and discusses their physical implications in the context of previously reported dark energy star models. Throughout this study, geometrized units are adopted such that $c = G = 1$.

\section{Interior Spacetime and Field Equations\label{sec:2}}

Einstein's field equations in the presence of a cosmological constant are given by \cite{1973grav.book.....M}
\begin{equation}
    \mathcal{R}_{\mu\nu} -\frac{1}{2}g_{\mu\nu}\mathcal{R} + g_{\mu\nu} \Lambda= 8\pi T_{\mu\nu},\label{Einstein_field_equations}
\end{equation}
where $\mathcal{R}_{\mu\nu}$ is the Ricci tensor, $\mathcal{R}$ is the Ricci scalar, $T_{\mu\nu}$ is the energy-momentum tensor, $g_{\mu\nu}$ is the metric tensor and $\Lambda$ is the cosmological constant. The general line element representing the interior solution of a static, spherically symmetric sphere in four--dimensional Schwarzschild coordinates can be written as
\begin{equation}
    ds^2 = -e^{\nu} dt^2 + e^{\lambda} dr^2 + r^2 d\theta^2 + r^2\sin^2{\theta}d\phi^2,
\end{equation}
where $e^{\nu} = g_{tt}$ and $e^{\lambda} = g_{rr}$ are metric potentials that depend on the radial coordinate $r$ only.

The nonzero components of the energy-momentum tensor can be expressed as
\begin{align}
    T_0^0 &\equiv \rho_{eff} = (\rho_{om} + \rho_{de}),\label{DF-SEMT-1}\\
    T_1^1 &\equiv -(p_r)_{eff} = -(p_{om} + (p_r)_{de}),\label{DF-SEMT-2}\\
    T_2^2 \equiv T_3^3 &\equiv -(p_t)_{eff} = -(p_{om} + (p_t)_{de}),\label{DF-SEMT-3}
\end{align}where $\rho$ denotes the energy density, $p_r$ represents the radial pressure, $p_t$ denotes the transverse pressure and the subscripts $de$ and $om$ refer to dark energy and ordinary matter, respectively.

The field equations in \autoref{Einstein_field_equations} can be reduced to
\begin{align}
    8\pi \rho_{eff} &= e^{-\lambda}\left(\frac{\lambda'}{r} - \frac{1}{r^2}\right) + \frac{1}{r^2} - \Lambda,\label{EFE-1}\\
    8\pi (p_r)_{eff} &= e^{-\lambda}\left(\frac{\nu'}{r} + \frac{1}{r^2}\right) - \frac{1}{r^2} + \Lambda,\label{EFE-2}\\
    8\pi (p_t)_{eff} &= \frac{e^{-\lambda}}{2}\left(\frac{\nu'^2-\lambda'\nu'}{2} + \frac{\nu'-\lambda'}{r} + \nu''\right) + \Lambda,\label{EFE-3}
\end{align}where $'$ denotes differentiation with respect to $r$. In this study, the metric potential $e^{\lambda}$ is chosen in the form
\begin{equation}\label{grr}
e^{\lambda(r)} = 1 + \frac{r^2}{R_*^2},
\end{equation}
where $R_*$ denotes the stellar radius. The remaining metric potential \( e^{\nu} \) is derived in the following section.

\section{The Equation of State and Stellar Modeling\label{sec:3}}

The temporal metric potential \( e^{\nu} \) can be derived using the {complexity factor formalism}, as introduced by Ref~\cite{2018PhRvD..97d4010H}, for static and self-gravitating fluid configurations. This formalism originates from the orthogonal splitting of the Riemann tensor and combines the effects of density inhomogeneity and pressure anisotropy into a single scalar quantity known as the complexity factor:
\begin{equation}\nonumber
    Y_{TF} = 8\Pi - \frac{4\pi}{R_*^3} \int_0^r \tilde{r}^3 \left( \rho_{{eff}} \right)' d\tilde{r},
\end{equation}
where \( \Pi = \tilde{p}_r - \tilde{p}_t \). Imposing a condition such as vanishing complexity (i.e., \( Y_{TF} = 0 \)) provides a physically motivated constraint, enabling the determination of the metric potential \( e^{\nu} \) in a non-arbitrary and structure-aware manner.

The metric potential $e^{\nu}$ is derived as shown by Refs. \cite{2024ChJPh..87..608R,2022EPJC...82..706C}, leading to the expression
\begin{align}
e^{\nu(r)} &= \left(A\int r e^{\lambda/2} dr +B \right)^2,\nonumber\\
&= \left(\frac{A \left(R_*^{2} + r^{2}\right)^{\frac{3}{2}}}{3 R_*} + B\right)^{2}\label{gtt},
\end{align}
where $A$ and $B$ are constants determined using the boundary conditions outlined in \autoref{sec:4}. The expression for \autoref{gtt} obtained in this work differs from the metric potential $e^{\nu}$ presented in Ref.~\cite{2024ChJPh..87..608R} by a factor of $R_* = 1/\sqrt{C}$.

The dark energy radial pressure is adopted as
\begin{equation}\label{eq:11}
    (p_r)_{de} = -\rho_{de},
\end{equation}
following Refs.~\cite{2021PDU....3400879B,2011Ap&SS.333..437G}, while the relationship between the ordinary matter density and the dark energy density is given by
\begin{equation}\label{de_om_energy_density_relation}
\rho_{de} = \alpha \rho_{om},
\end{equation}
where \( \alpha \) is a nonzero coupling parameter. The values of the coupling parameter were investigated in Ref.~\cite{2024ChJPh..92.1703A}, where $\alpha \rightarrow 1$ corresponds to an equal proportion of ordinary matter and dark energy, while $\alpha \rightarrow \infty$ corresponds to a higher contribution of dark energy. Throughout this study, we assume a balanced contribution of dark energy and ordinary matter by adopting $\alpha = 1$.

Using \autoref{DF-SEMT-1}--\autoref{de_om_energy_density_relation}, the expressions for the effective energy density, effective radial pressure and effective transverse pressure are obtained as follows
\begin{equation}
    \rho_{eff}{\left(r \right)} = \frac{
3R_*^{2} + r^2 - \Lambda (R_*^{2}+r^2)^2
}{
8\pi (R_*^{2}+r^2)^2
},
\end{equation}

\begin{equation}\label{eq:pr_eff}
    ({p_r})_{eff}{\left(r \right)} = \frac{
A\sqrt{R_*^{2}+r^2}\left[\Lambda (R_*^{2}+r^2)^2 + 5R_*^{2} - r^2\right]
+ 3BR_*\left[\Lambda (R_*^{2}+r^2)-1\right]
}{
8\pi (R_*^{2}+r^2)\left[
A(R_*^{2}+r^2)^{3/2}+3BR_*
\right]
},
\end{equation}

\begin{equation}\label{eq:pt_eff}
    {(p_t)}_{eff}{\left(r \right)} = \frac{
A^2 (R_*^{2}+r^2)^4 \left[\Lambda (R_*^{2}+r^2)+10\right]
+ 6ABR_*\sqrt{R_*^{2}+r^2}\,(R_*^{2}+r^2)^2
\left[\Lambda (R_*^{2}+r^2)+4\right]
+ 9B^2R_*^{2}\left[\Lambda (R_*^{2}+r^2)^2-2R_*^{2}\right]
}{
16\pi (R_*^{2}+r^2)^2
\left[
A(R_*^{2}+r^2)^{3/2}+3BR_*
\right]^2
}.
\end{equation}
At the stellar center $(r=0)$, the effective energy density, radial pressure and transverse pressure reduce to
\begin{equation}
    \rho_{eff}(0)
    =
    \frac{3-\Lambda R_*^{2}}
    {8\pi R_*^{2}},
\end{equation}

\begin{equation}
    (p_r)_{eff}(0)
    =
    \frac{
    A\left(\Lambda R_*^{2}+5\right)
    +
    3B\left(\Lambda R_*^{2}-1\right)
    }
    {
    8\pi R_*^{2}(A+3B)
    },
\end{equation}

\begin{equation}
    (p_t)_{eff}(0)
    =
    \frac{
    A^{2}\left(\Lambda R_*^{2}+10\right)
    +
    6AB\left(\Lambda R_*^{2}+4\right)
    +
    9B^{2}\left(\Lambda R_*^{2}-2\right)
    }
    {
    16\pi R_*^{2}(A+3B)^{2}
    }.
\end{equation}
Further, subtracting \autoref{eq:pr_eff} from \autoref{eq:pt_eff} yields the pressure anisotropy, defined as \( \Delta = (p_t)_{{eff}} - (p_r)_{{eff}} \), given by
\begin{equation}
    \Delta(r) = \frac{- \Lambda R_*^{4} - 2 \Lambda R_*^{2} r^{2} - \Lambda r^{4} + 2 r^{2}}{16 \pi \left(R_*^{4} + 2 R_*^{2} r^{2} + r^{4}\right)}.
\end{equation}
All thermodynamic quantities remain finite and regular throughout the configuration. Moreover, evaluating the anisotropy at the stellar center yields
\[
\Delta(0)= -\frac{\Lambda}{16\pi}.
\]
Therefore, the anisotropy vanishes only for the Schwarzschild case \((\Lambda=0)\), whereas nonzero cosmological constants introduce a finite central anisotropy. Unlike conventional anisotropic compact star models, this central anisotropy originates directly from the cosmological constant contribution.

\section{Exterior Spacetime and Boundary Conditions \label{sec:4}}

The interior spacetime is smoothly matched to the exterior Schwarzschild--(anti--)de Sitter geometry at the stellar boundary \( r=R_* \). The corresponding exterior line element is given by
\begin{equation}\label{line_element_Sch}
    ds^2 = -\left(1 - \frac{2M}{r} - \frac{\Lambda r^2}{3}\right) dt^2
    + \left(1 - \frac{2M}{r} - \frac{\Lambda r^2}{3}\right)^{-1}dr^2
    + r^2 d\theta^2 + r^2 \sin^2\theta \, d\phi^2 .
\end{equation}

A physically viable compact configuration requires the continuity of the metric components \( g_{tt} \) and \( g_{rr} \), together with the continuity of \( \partial g_{tt}/\partial r \), across the hypersurface \( r=R_* \). Accordingly, the matching conditions are expressed as
\begin{equation}\label{equalize-grr}
    g_{rr} : \left(1+\frac{R_*^2}{R_*^{2}}\right)
    =
    \left(1-\frac{2M}{R_*}-\frac{\Lambda R_*^2}{3}\right)^{-1}.
\end{equation}
Solving \autoref{equalize-grr}, one obtains
\begin{equation}
    2\Lambda R_*^3 - 3R_* + 12M = 0.
\end{equation}
This relation determines the stellar radius \(R_*\) for a given mass \(M\) and cosmological constant \(\Lambda\) by selecting only the roots that are both real and positive. To perform the numerical analysis of the model, we consider the compact star candidate \textit{Vela X-1}, which has an observed mass 
\(
M = 1.77\,M_{\odot}
\)
\cite{2013MNRAS.431.3216G}. 
This object has been widely employed in compact star modeling and provides a suitable astrophysical framework for investigating the influence of the cosmological constant on dark energy star configurations.

The continuity condition for the temporal metric component leads to
\begin{equation}\label{lambda_lambda}
    g_{tt} : \left(
    \frac{A\left(2R_*^2\right)^{3/2}}{3R_*} + B
    \right)^2
    =
    \left(
    1-\frac{2M}{R_*}-\frac{\Lambda R_*^2}{3}
    \right),
\end{equation}
whereas continuity of its radial derivative yields
\begin{equation}\label{nu_nu}
    \partial g_{tt}/\partial r: 2A R_* \sqrt{2}
    \left(
    \frac{2A\sqrt{2}R_*^2}{3} + B
    \right)
    =
    \frac{2M}{R_*^2}
    - \frac{2\Lambda R_*}{3}.
\end{equation}
Simultaneously solving \autoref{lambda_lambda} and \autoref{nu_nu} gives
\begin{equation}\label{constant_A}
A
=
\frac{
3M-\Lambda R_*^3
}{
3\sqrt{2}\,R_*^3
\sqrt{
1-\dfrac{2M}{R_*}-\dfrac{\Lambda R_*^2}{3}
}
},
\end{equation}
and
\begin{equation}\label{constant_B}
B
=
\frac{
9R_*-24M-\Lambda R_*^3
}{
9R_*
\sqrt{
1-\dfrac{2M}{R_*}-\dfrac{\Lambda R_*^2}{3}
}
}.
\end{equation}

The constants \(A\) and \(B\) depend explicitly on the stellar mass \(M\), radius \(R_*\) and cosmological constant \(\Lambda\). In order to investigate the influence of the cosmological constant on the internal structure and physical behavior of dark energy stars, both positive and negative values of \(\Lambda\), including the limiting case \(\Lambda=0\) (Schwarzschild solution), are considered. This choice is motivated by investigations of Schwarzschild--de Sitter ($\Lambda > 0$) and Schwarzschild--anti--de Sitter ($\Lambda < 0$) geometries, where the cosmological constant plays an important role in modifying the horizon structure, geodesic motion and tidal properties of compact gravitational systems \cite{2021EPJC...81..610V}.

The values \(\Lambda=\pm10^{-5}\) and \(\Lambda=\pm10^{-3}\) are adopted phenomenologically to represent weak and comparatively enhanced cosmological constant contributions within the stellar interior. Although these values are considerably larger than the observed cosmological background value, \(\Lambda\sim10^{-46}\,\mathrm{km}^{-2}\) \cite{2020A&A...641A...6P,2021A&A...652C...4P}, similar effective local cosmological constants have been employed in relativistic compact-star models to explore the role of cosmological constant in dense gravitational systems \cite{2025AnPhy.47870029S}. The adopted values should therefore be interpreted as effective local cosmological contributions within compact gravitational systems rather than as the cosmological background value itself. The numerical values of the model parameters and the corresponding stellar radii obtained for different values of the cosmological constant are summarized in \autoref{tab:lambda_variation}, which shows that the stellar radius increases for positive values of \(\Lambda\) and decreases for negative values of \(\Lambda\).

\begin{table}[!ht]
\centering
\caption{
Values of the model parameters $A$, $B$ and stellar radius $R_*$ obtained for different values of the cosmological constant $\Lambda$ in the compact star Vela X-1 with mass $M = 1.77\,M_{\odot}$. Here, SdS and SAdS denote the Schwarzschild--de Sitter and Schwarzschild--anti--de Sitter geometries, respectively.
}
\scriptsize
\begin{tabular}{lcccc}
\toprule
Geometry 
& $\Lambda$ 
& $A$ 
& $B$ 
& $R_*$
\\

& (km$^{-2}$)
& (km$^{-2}$)
& 
& (km)
\\
\midrule

SdS
& $10^{-3}$
& 0.00140
& 0.53310
& 11.45
\\

SdS
& $10^{-5}$
& 0.00227
& 0.47192
& 10.46
\\

Schwarzschild
& $0$
& 0.00228
& 0.47140
& 10.45
\\

SAdS
& $-10^{-5}$
& 0.00229
& 0.47088
& 10.44
\\

SAdS
& $-10^{-3}$
& 0.00309
& 0.42592
& 9.82
\\

\bottomrule
\end{tabular}
\label{tab:lambda_variation}
\end{table}

\section{Physical Attributes and Stability Analysis\label{sec:5}}

\subsection{Spacetime Geometry and Regularity}

To verify the physical acceptability of our dark energy star model, we first examine the behavior of the spatial and temporal metric potentials $e^{\lambda}$ and $e^{\nu}$.

\begin{figure}[htbp]
    \centering
    \includegraphics[width=0.47\linewidth]{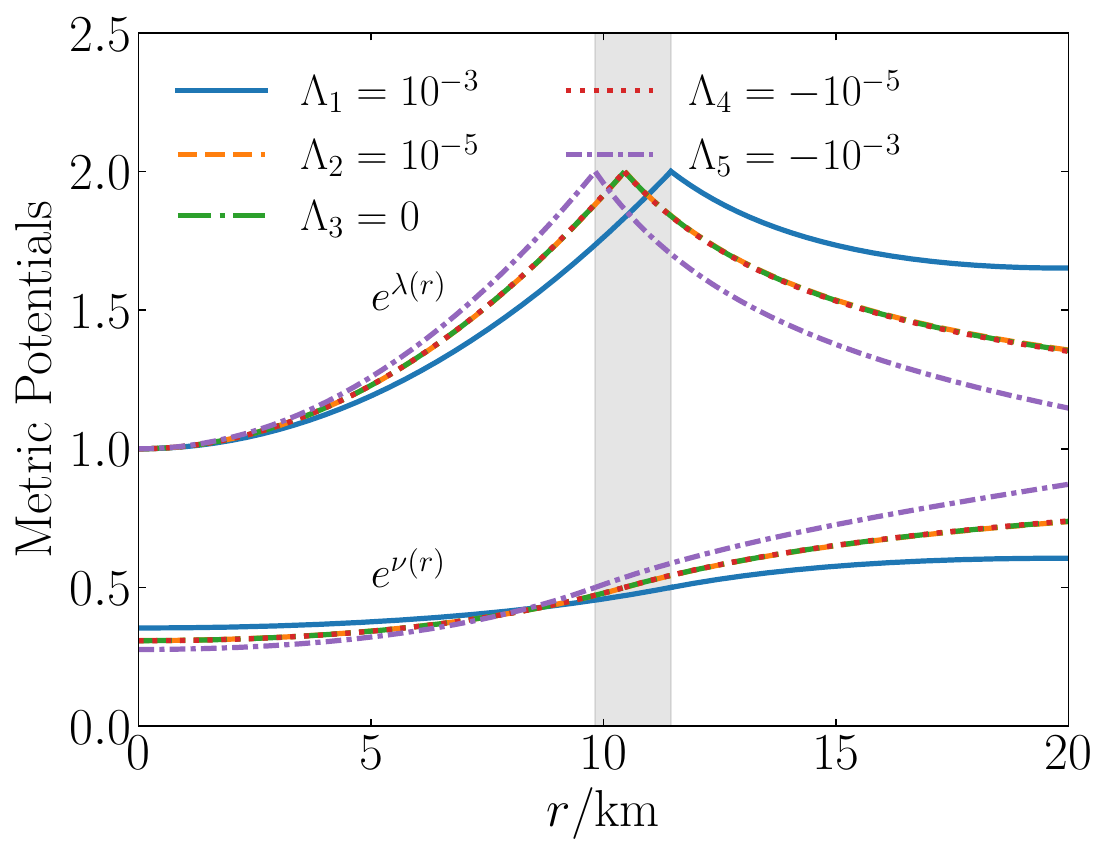}
    \caption{Radial variation of the metric potentials $e^{\lambda(r)}$ and $e^{\nu(r)}$ for different values of the cosmological constant $\Lambda = 10^{-3}, 10^{-5}, 0, -10^{-5}$ and $-10^{-3}\,\mathrm{km}^{-2}$. The shaded region represents the matching region between the interior and exterior spacetimes.}
    \label{fig:metric}
\end{figure}As depicted in \autoref{fig:metric}, both metric potentials are finite, positive and regular everywhere inside the stellar interior, monotonically increasing from the core to the surface. At the exact center of the star ($r = 0$), we observe that $e^{\lambda(0)}$ is equal to a non-zero positive constant. Furthermore, the first derivatives of these metric potentials vanish at the origin, confirming that the spacetime geometry is completely free of any physical or geometric singularities at the core.

\subsection{Thermodynamic Matter Profiles and Anisotropy}

\begin{figure}[htbp]
    \centering

    \begin{subfigure}{0.47\textwidth}
        \centering
        \includegraphics[width=\linewidth]{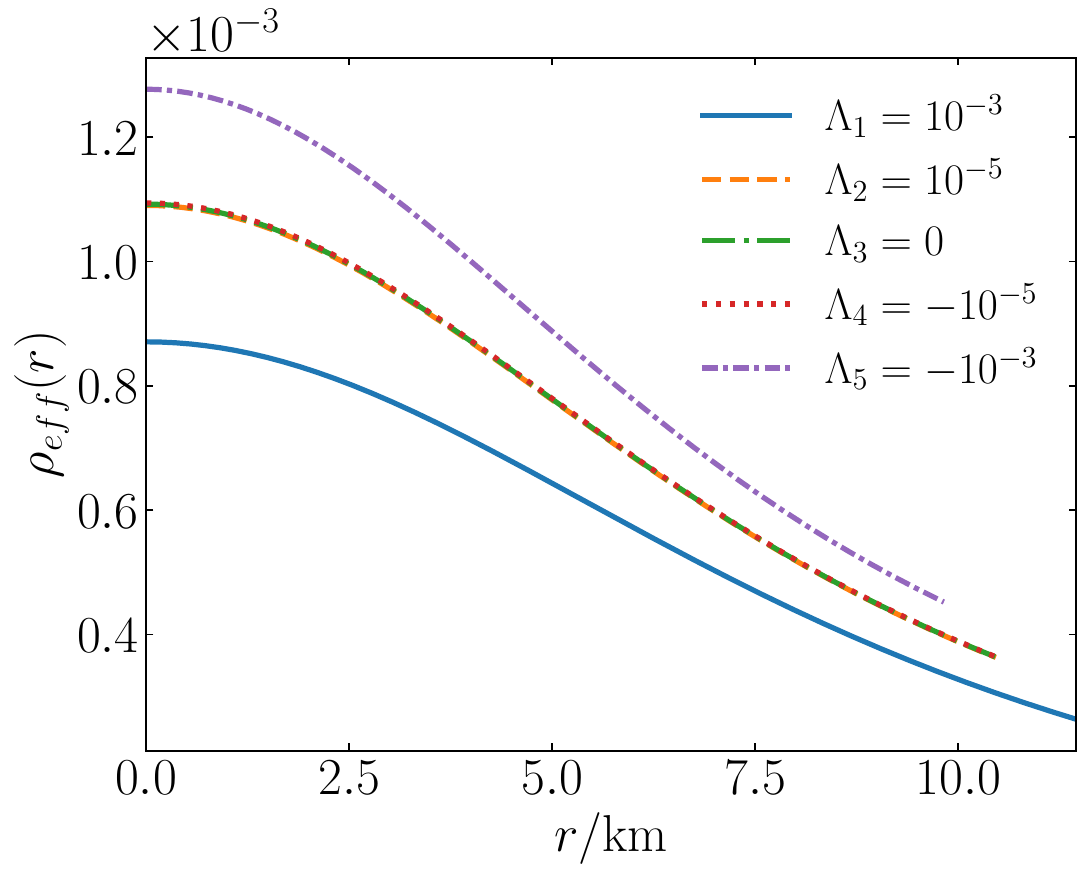}
        \caption{Effective energy density $\rho_{eff}(r)$.}
        \label{fig:density}
    \end{subfigure}
    \hfill
    \begin{subfigure}{0.47\textwidth}
        \centering
        \includegraphics[width=\linewidth]{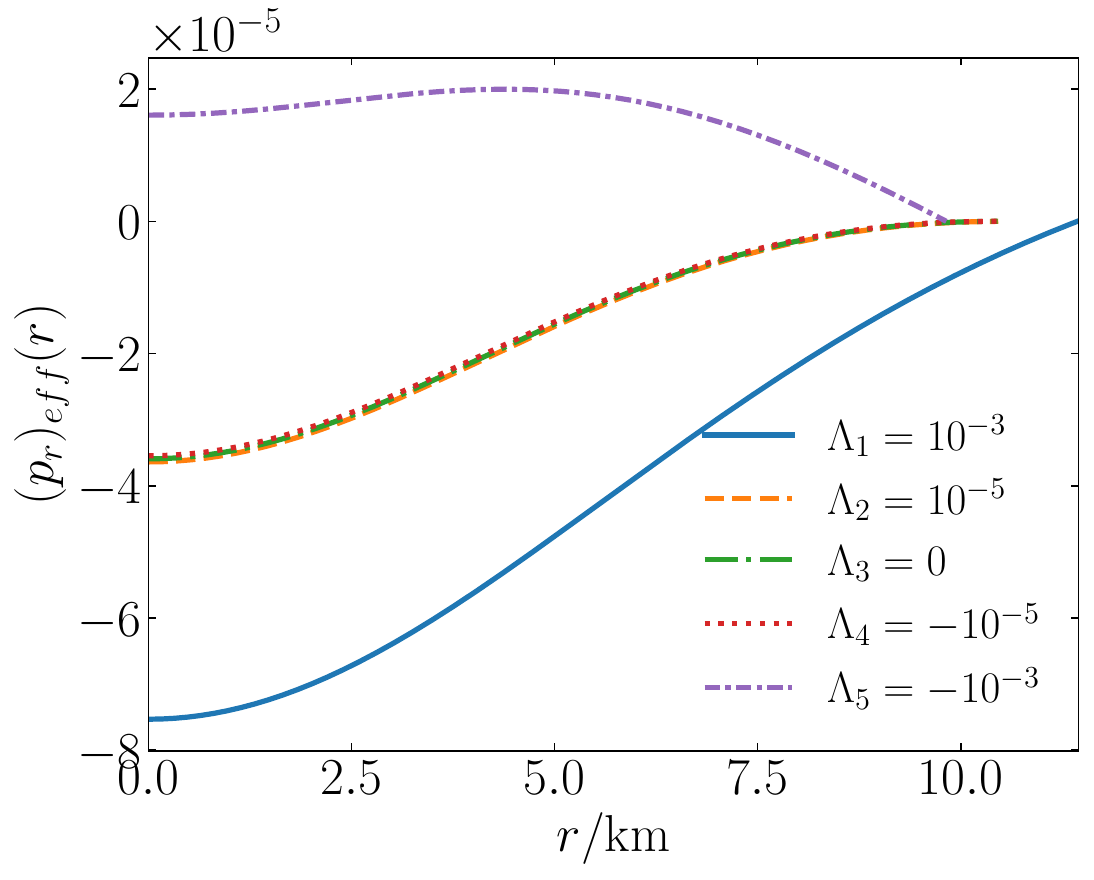}
        \caption{Effective radial pressure $(p_r)_{eff}(r)$.}
        \label{fig:radial_pressure}
    \end{subfigure}

    \vspace{0.5cm}

    \begin{subfigure}{0.47\textwidth}
        \centering
        \includegraphics[width=\linewidth]{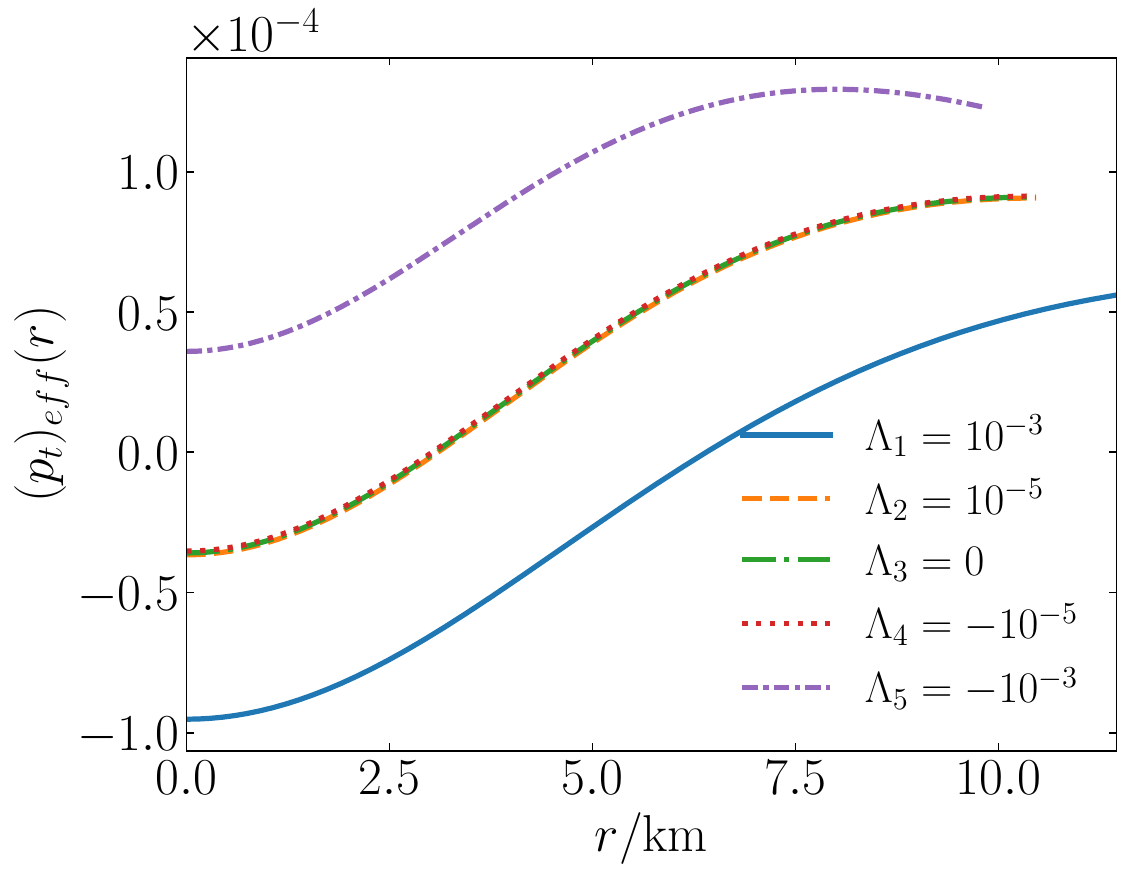}
        \caption{Effective transverse pressure $(p_t)_{eff}(r)$.}
        \label{fig:transverse_pressure}
    \end{subfigure}
    \hfill
    \begin{subfigure}{0.47\textwidth}
        \centering
        \includegraphics[width=\linewidth]{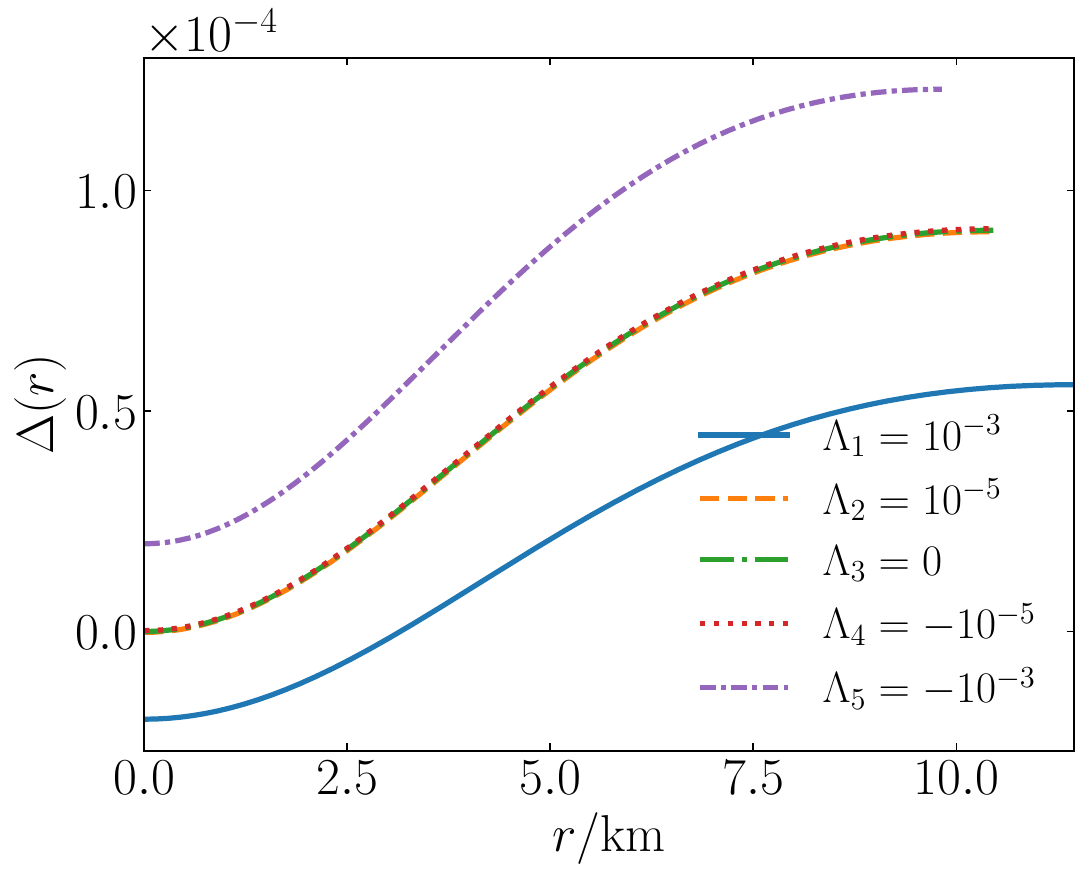}
       \caption{Pressure anisotropy $\Delta(r)$.}
        \label{fig:anisotropy}
    \end{subfigure}

    \caption{Radial behavior of the thermodynamic matter profiles and anisotropy for different values of the cosmological constant $\Lambda = 10^{-3}, 10^{-5}, 0, -10^{-5}$ and $-10^{-3}\,\mathrm{km}^{-2}$.}
    \label{fig:combined_thermodynamic_profiles}
\end{figure}

The radial variations of the effective energy density $\rho$, radial pressure $p_r$, transverse pressure $p_t$ and anisotropy $\Delta$ are illustrated in \autoref{fig:combined_thermodynamic_profiles}.

The energy density shown in \autoref{fig:density} attains its maximum value at the stellar core, indicating a highly dense central region and decreases monotonically toward the surface. Importantly, the radial pressure shown in \autoref{fig:radial_pressure} vanishes at the stellar boundary (\(r=R\)), decreasing for \(\Lambda = -10^{-3}\) and increasing for the remaining values of \(\Lambda\), thereby physically defining the surface of the star. The behavior of the transverse pressure is presented in \autoref{fig:transverse_pressure}, showing a trend similar to that of the radial pressure but with slightly larger values throughout the stellar interior. The anisotropic factor, which measures the deviation from isotropic pressure, is shown in \autoref{fig:anisotropy}. The anisotropy parameter remains positive throughout the stellar interior for all considered values of the cosmological constant, indicating that the anisotropic force is outwardly directed and repulsive in nature. Furthermore, the magnitude of the anisotropy increases with radial distance from the center. For sufficiently small values of the cosmological constant, the anisotropy approaches zero near the center, whereas larger values of positive or negative \(\Lambda\) produce small but non-vanishing central anisotropy.

\subsection{Macroscopic Observables}

\begin{figure}[htbp]
    \centering

    \begin{subfigure}{0.47\textwidth}
        \centering
        \includegraphics[width=\linewidth]{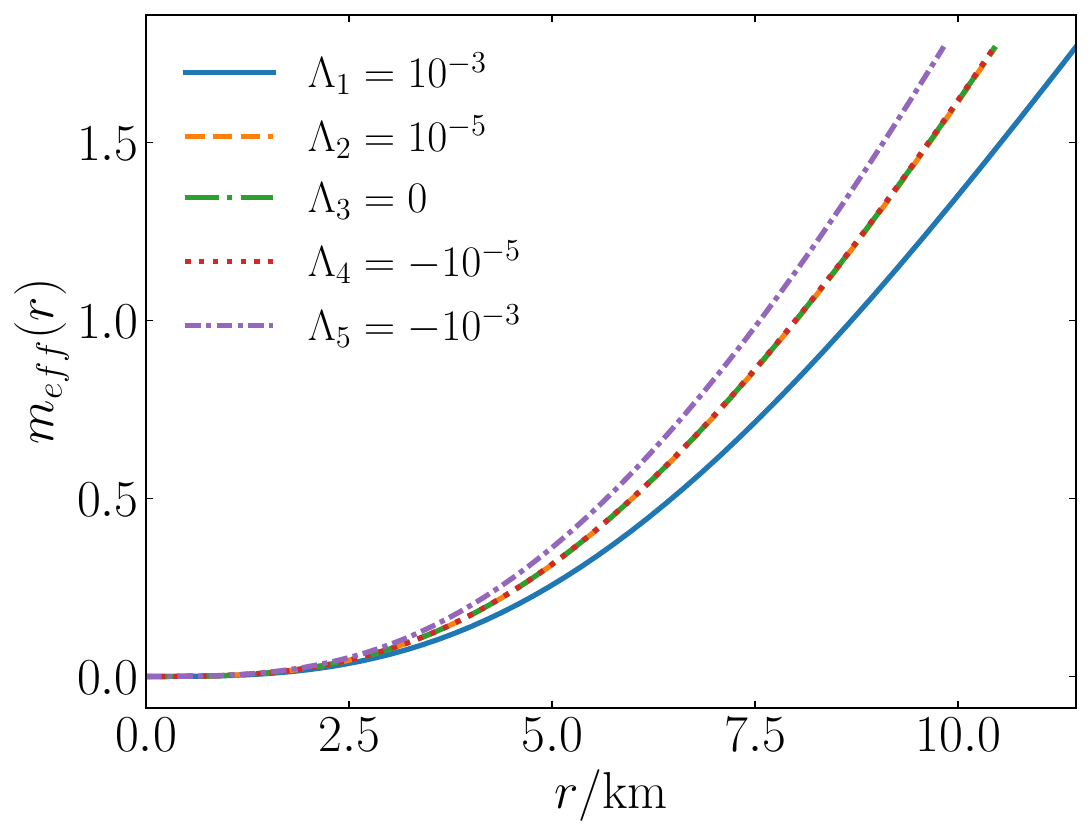}
        \caption{Effective mass function $m_{eff}(r)$.}
        \label{fig:mass}
    \end{subfigure}
    \hfill
    \begin{subfigure}{0.47\textwidth}
        \centering
        \includegraphics[width=\linewidth]{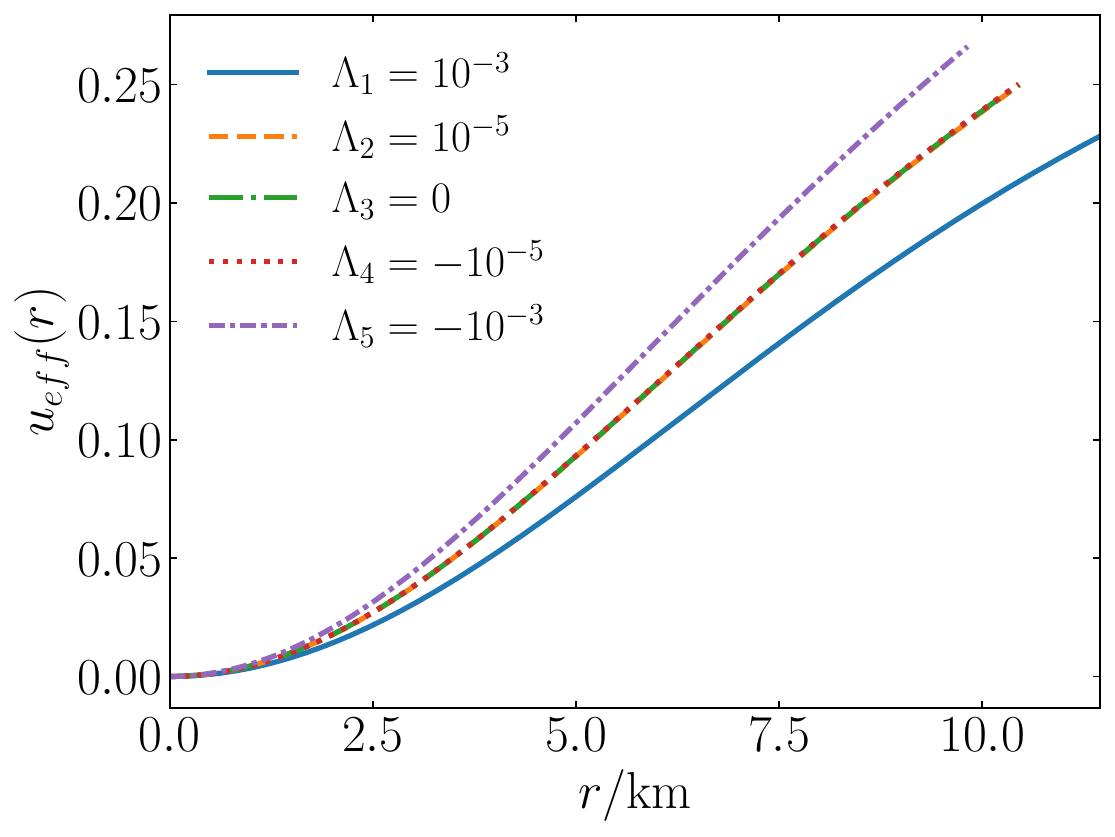}
        \caption{Effective compactness $u_{eff}(r)$.}
        \label{fig:compactness}
    \end{subfigure}

    \vspace{0.5cm}

    \begin{subfigure}{0.47\textwidth}
        \centering
        \includegraphics[width=\linewidth]{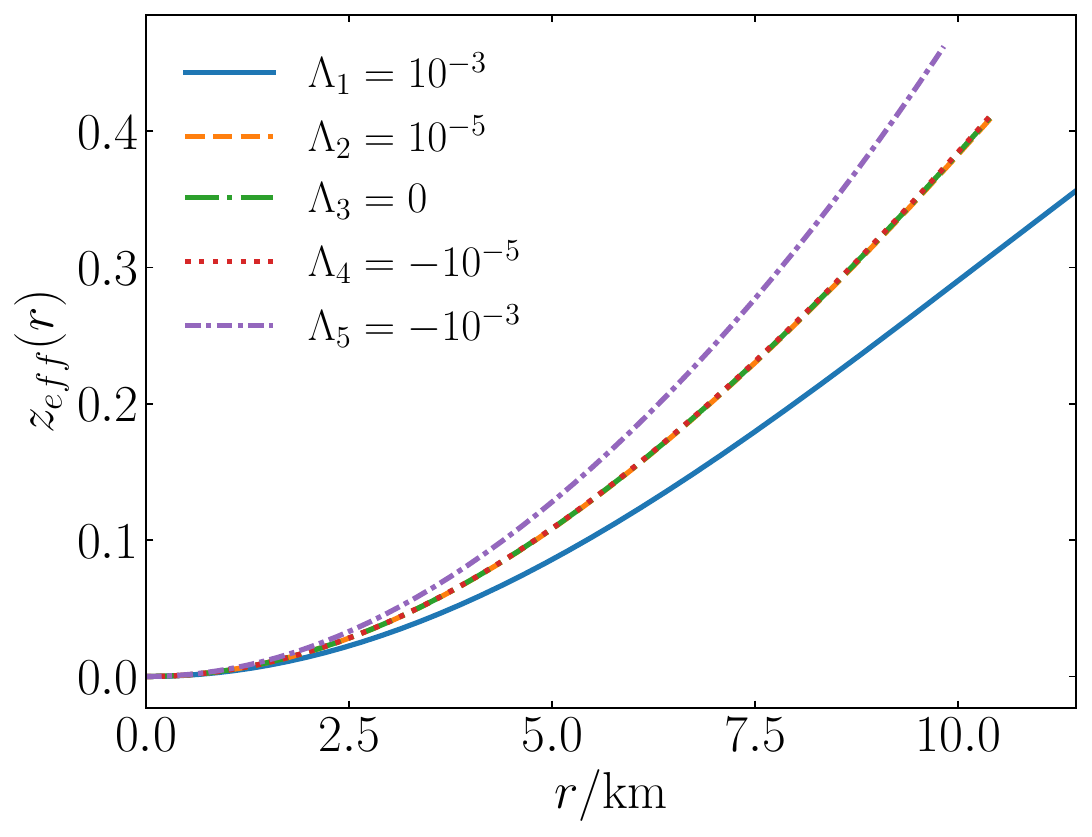}
        \caption{Effective redshift function $z_{eff}(r)$.}
        \label{fig:redshift}
    \end{subfigure}

    \caption{Radial profiles of the effective mass $m_{eff}(r)$, effective compactness $u_{eff}(r)$ and effective redshift function $z_{eff}(r)$ for different values of the cosmological constant $\Lambda = 10^{-3}, 10^{-5}, 0, -10^{-5}$ and $-10^{-3}\,\mathrm{km}^{-2}$.}
    \label{fig:massss}
\end{figure}

In this section, the equations and results related to key parameters such as the mass function, compactness and redshift function are presented. By equating \autoref{grr} with the radial component of the exterior solution given in \autoref{line_element_Sch} and solving for \( m(r) \), the effective mass function is obtained as
\begin{equation}\label{eq:mass_function_eff}
    m_{{eff}}(r) = \frac{
r^3 \left[3-\Lambda (R_*^2+r^2)\right]
}{
6(R_*^2+r^2)
}.
\end{equation}\autoref{fig:mass} presents the effective mass function, which exhibits a regular, monotonically increasing profile from the center ($m(0)=0$) and reaches its maximum value at the stellar boundary to define the total mass $M$.

Dividing \autoref{eq:mass_function_eff} by \( r \) yields the expression for the effective compactness as
\begin{equation}\label{eq:compactness}
    u_{{eff}}(r) = \frac{
r^2 \left[3-\Lambda (R_*^2+r^2)\right]
}{
6(R_*^2+r^2)
}.
\end{equation}The results show that the maximum compactness is greatest for \(\Lambda = -10^{-3}\) and smallest for \(\Lambda = 10^{-3}\), as illustrated in \autoref{fig:compactness}. The compactness profiles corresponding to the remaining values of \(\Lambda\) are approximately identical.

Based on \autoref{eq:compactness}, the effective redshift function is obtained as
\begin{align}\label{eq:effective_redshift_function}
    z_{eff}{\left(r \right)} &= (1 - 2u(r))^{-1/2} - 1,\nonumber\\ &= \sqrt{
\frac{3(R_*^2+r^2)}
{\Lambda r^2 (R_*^2+r^2)+3R_*^2}
}
-1.
\end{align}The effective redshift function shown in \autoref{fig:redshift} remains positive, finite and increases toward the stellar surface. A behavior similar to that of the compactness is observed, where the surface redshift is largest for \(\Lambda = -10^{-3}\) and smallest for \(\Lambda = 10^{-3}\). The surface redshift corresponding to the remaining values of \(\Lambda\) is approximately identical.

\subsection{Verification of the Energy Conditions}

To validate the physical nature of the internal fluid, we evaluate the Null (NEC), Weak (WEC), Strong (SEC) and Dominant (DEC) energy conditions across the stellar interior. These conditions can be formulated generally as\begin{multicols}{2}\begin{enumerate}
  \item [(i)]WEC: $\rho + p_i \geq 0$, $\rho \geq 0$,
  \item [(ii)]NEC: $\rho + p_i \geq 0$,
  \item [(iii)]DEC: $\rho \geq |p_i|$, $\rho \geq 0$,
  \item [(iv)]SEC: $\rho + p_i \geq 0$, $\rho + p_r + 2p_t \geq 0$.
\end{enumerate}
\end{multicols}

\begin{figure}[!htbp]
    \centering

    \begin{subfigure}{0.47\textwidth}
        \centering
        \includegraphics[width=\linewidth]{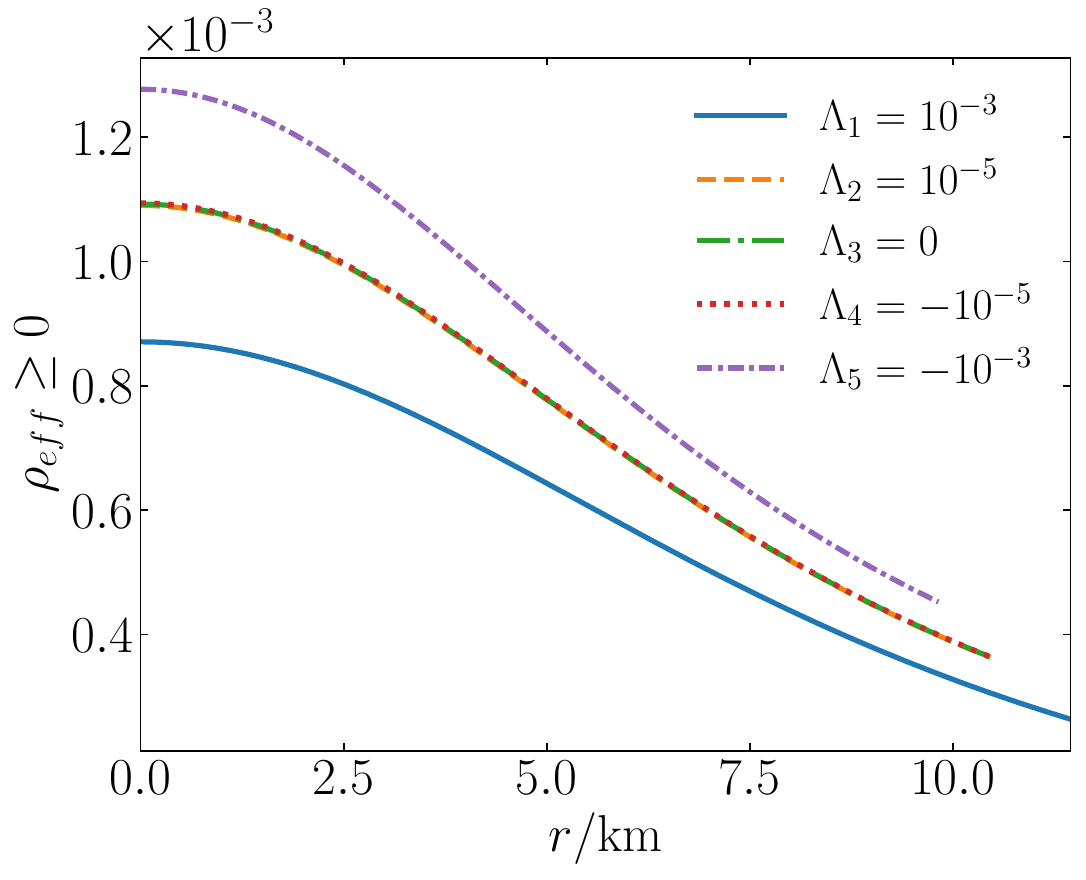}
        \caption{Weak energy condition $\rho \geq 0$.}
        \label{fig:WEC}
    \end{subfigure}
    \hfill
    \begin{subfigure}{0.47\textwidth}
        \centering
        \includegraphics[width=\linewidth]{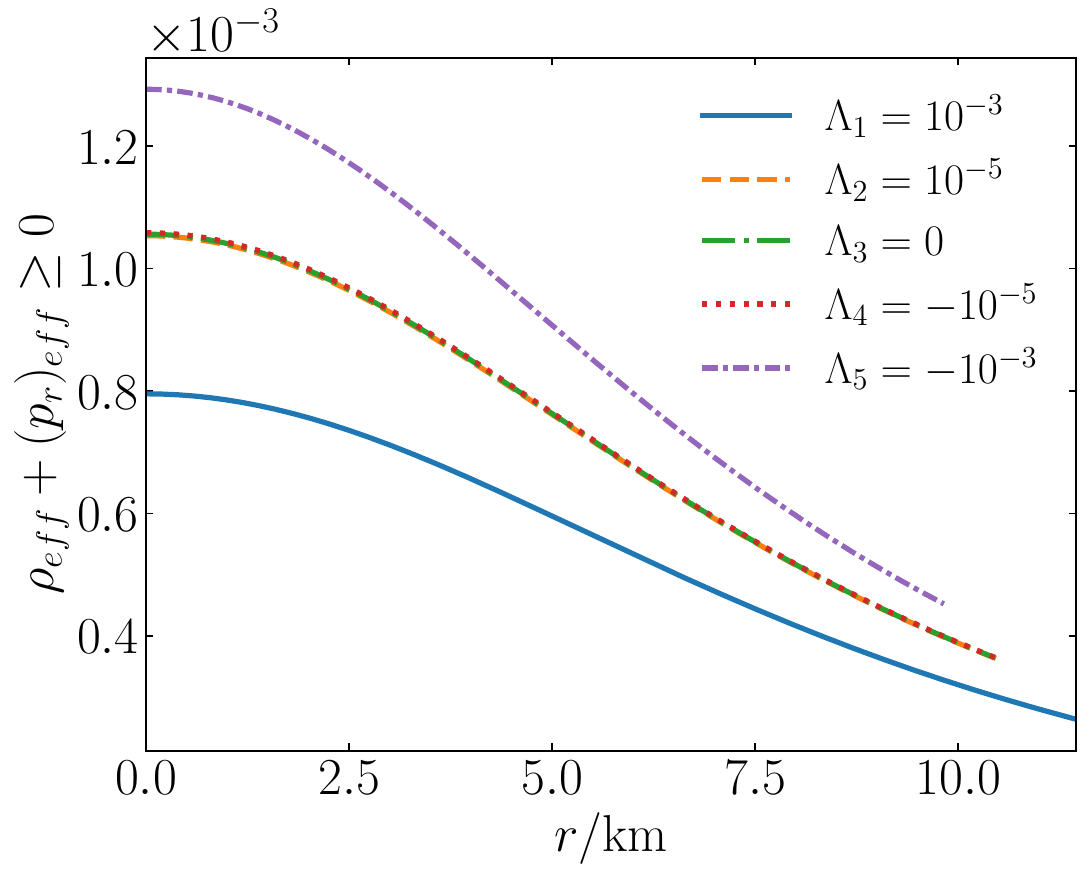}
        \caption{Null energy condition $\rho + p_r \geq 0$.}
        \label{fig:NECr}
    \end{subfigure}

    \vspace{0.2cm}

    \begin{subfigure}{0.47\textwidth}
        \centering
        \includegraphics[width=\linewidth]{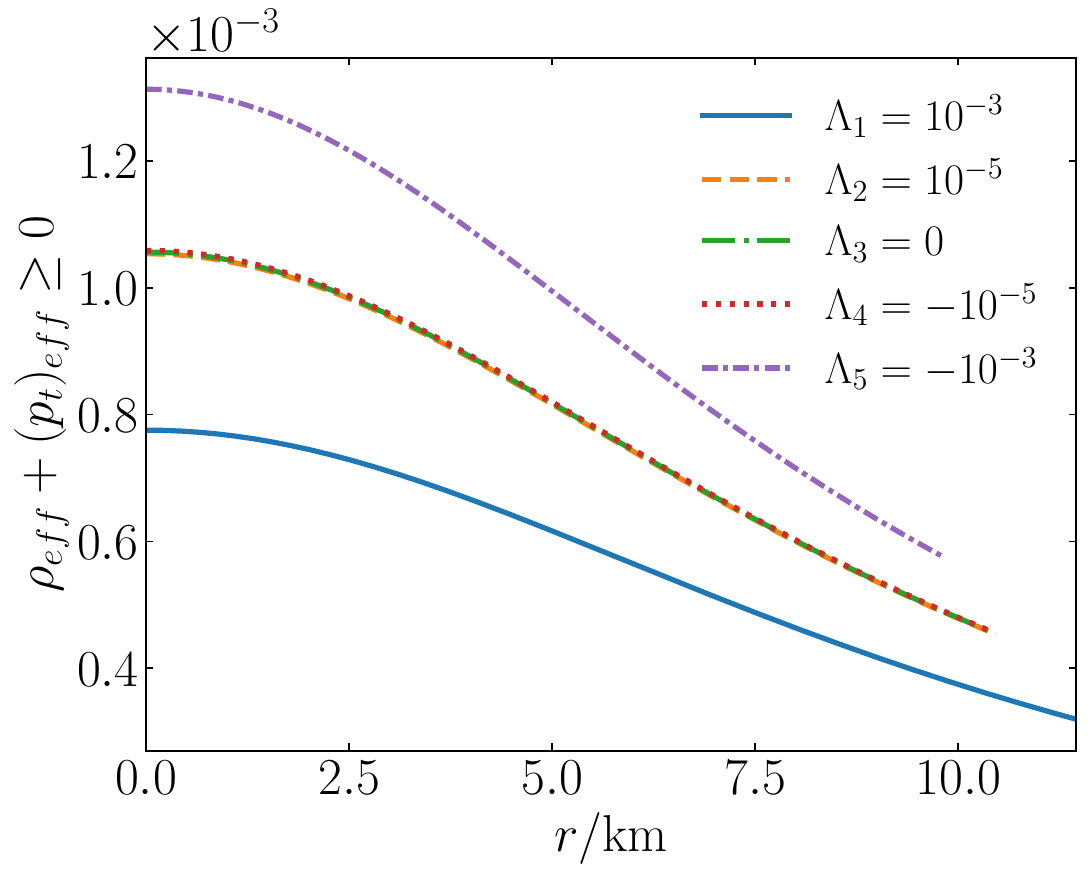}
        \caption{Null energy condition $\rho + p_t \geq 0$.}
        \label{fig:NECt}
    \end{subfigure}
    \hfill
    \begin{subfigure}{0.47\textwidth}
        \centering
        \includegraphics[width=\linewidth]{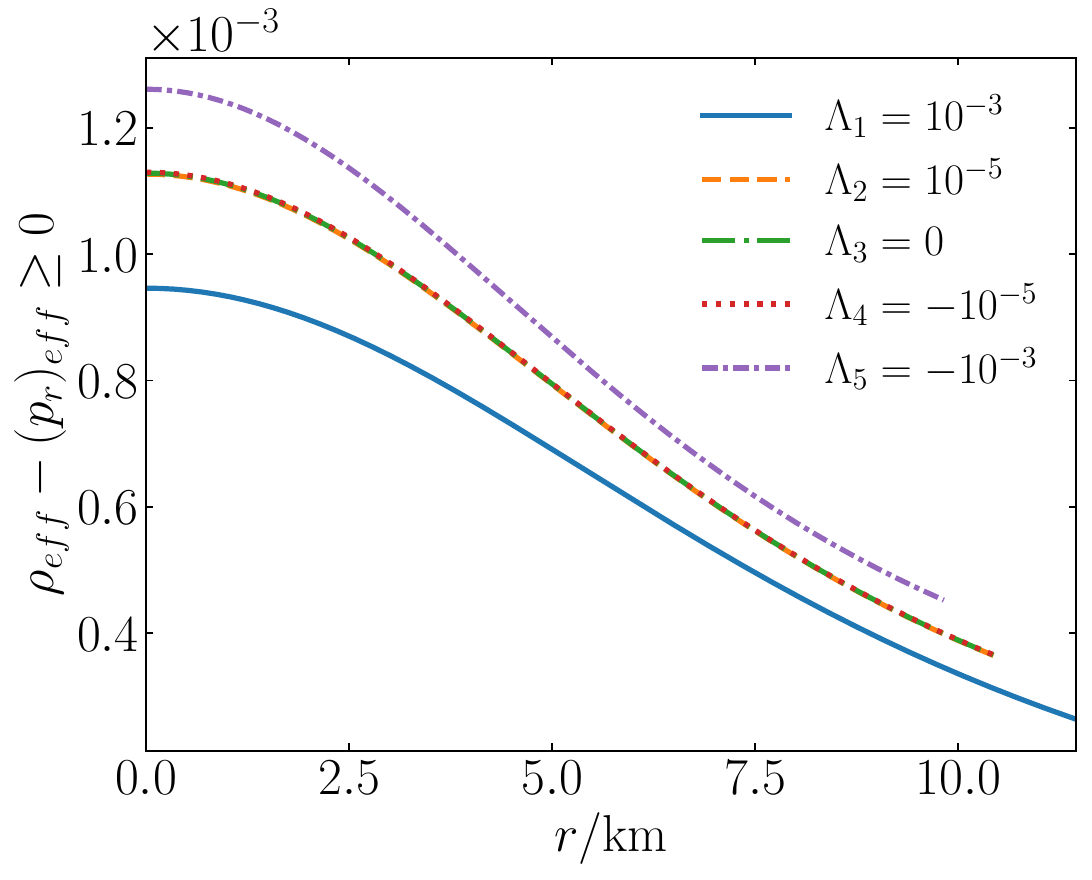}
        \caption{Dominant energy condition $\rho - p_r \geq 0$.}
        \label{fig:DECr}
    \end{subfigure}

    \vspace{0.2cm}

    \begin{subfigure}{0.47\textwidth}
        \centering
        \includegraphics[width=\linewidth]{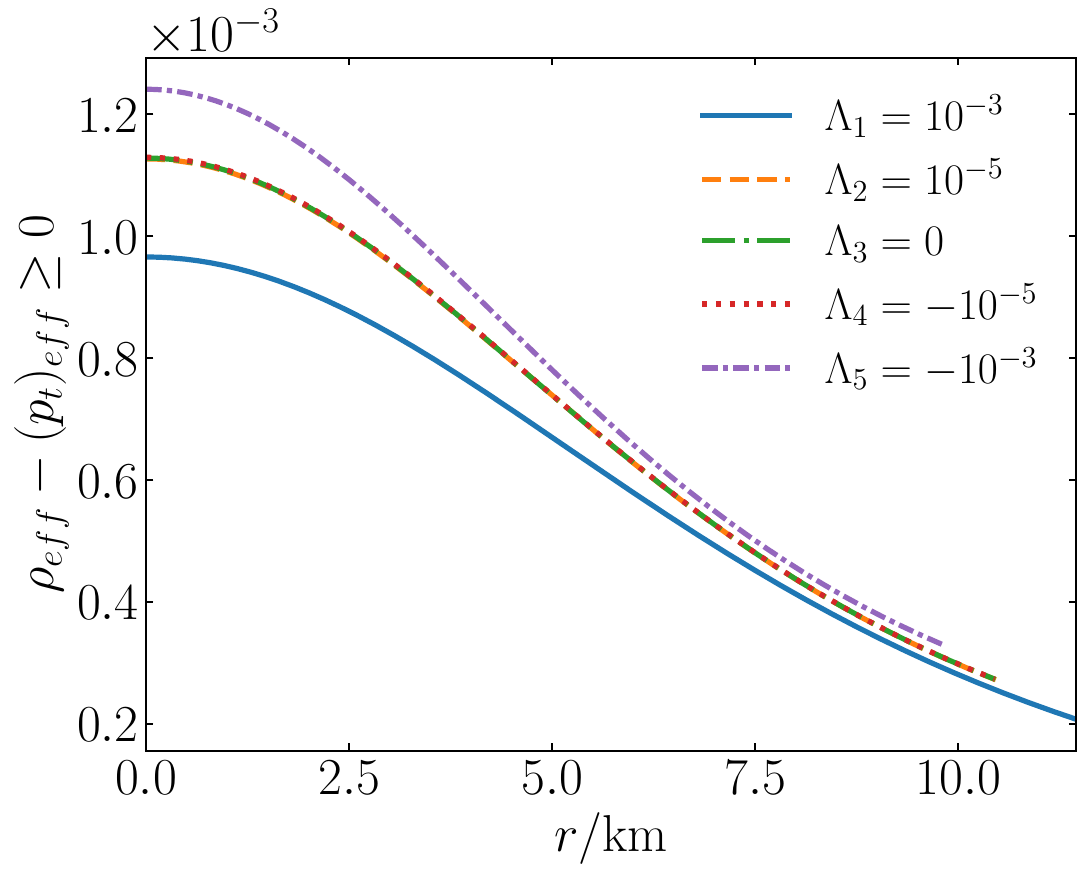}
        \caption{Dominant energy condition $\rho - p_t \geq 0$.}
        \label{fig:DECt}
    \end{subfigure}
    \hfill
    \begin{subfigure}{0.47\textwidth}
        \centering
        \includegraphics[width=\linewidth]{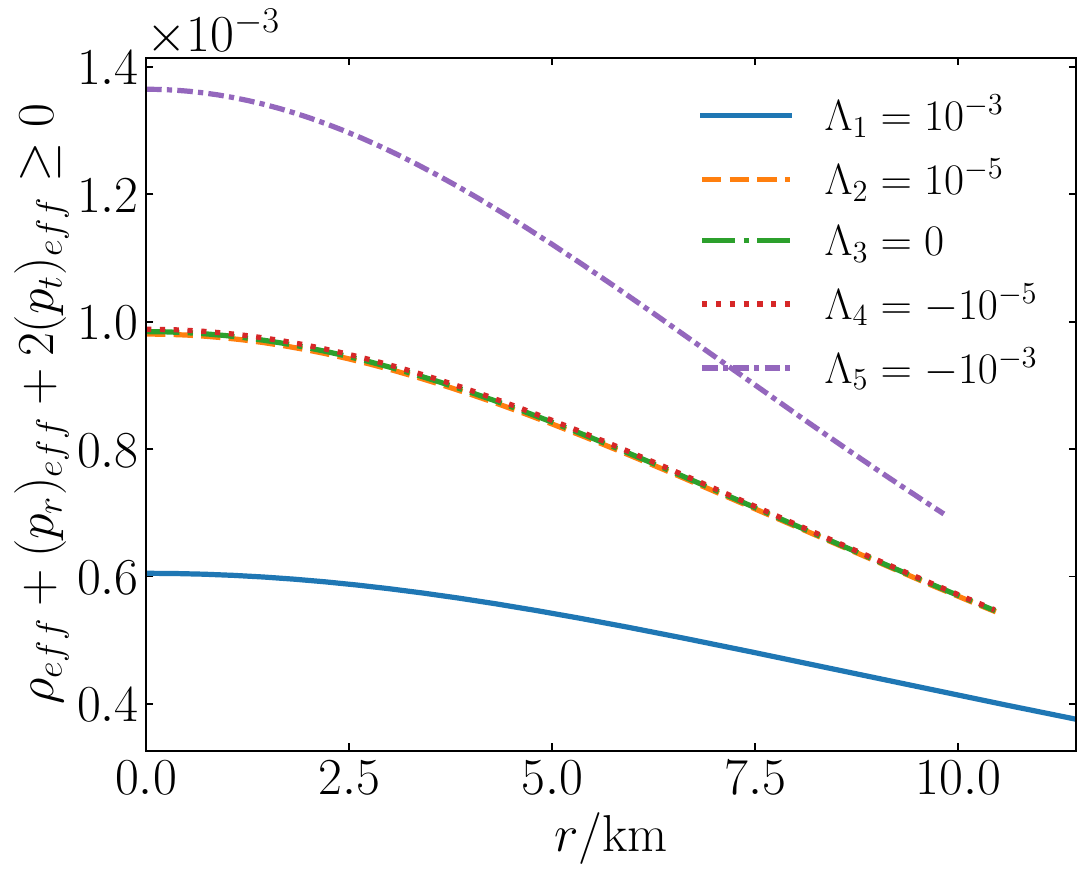}
        \caption{Strong energy condition $\rho + p_r + 2p_t \geq 0$.}
        \label{fig:SEC}
    \end{subfigure}

    \caption{Radial profiles of the energy conditions for different values of the cosmological constant $\Lambda = 10^{-3}, 10^{-5}, 0, -10^{-5}$ and $-10^{-3}\,\mathrm{km}^{-2}$.}
    \label{fig:energy-conditions}
\end{figure}

As illustrated in \autoref{fig:energy-conditions}, all considered energy conditions, namely the WEC, NEC, DEC and SEC are satisfied throughout the entire stellar interior for every considered value of the cosmological constant. In particular, both the radial and tangential forms of the null and dominant energy conditions remain positive everywhere, indicating that the effective energy density is non--negative and the matter distribution remains physically acceptable. Furthermore, the SEC is also satisfied throughout the configuration, demonstrating that the present model does not require any explicit violation of the strong energy condition to maintain equilibrium and stability.

\subsection{Hydrostatic Equilibrium}

\begin{figure}[htbp]
    \centering

    \begin{subfigure}{0.47\textwidth}
        \centering
        \includegraphics[width=\linewidth]{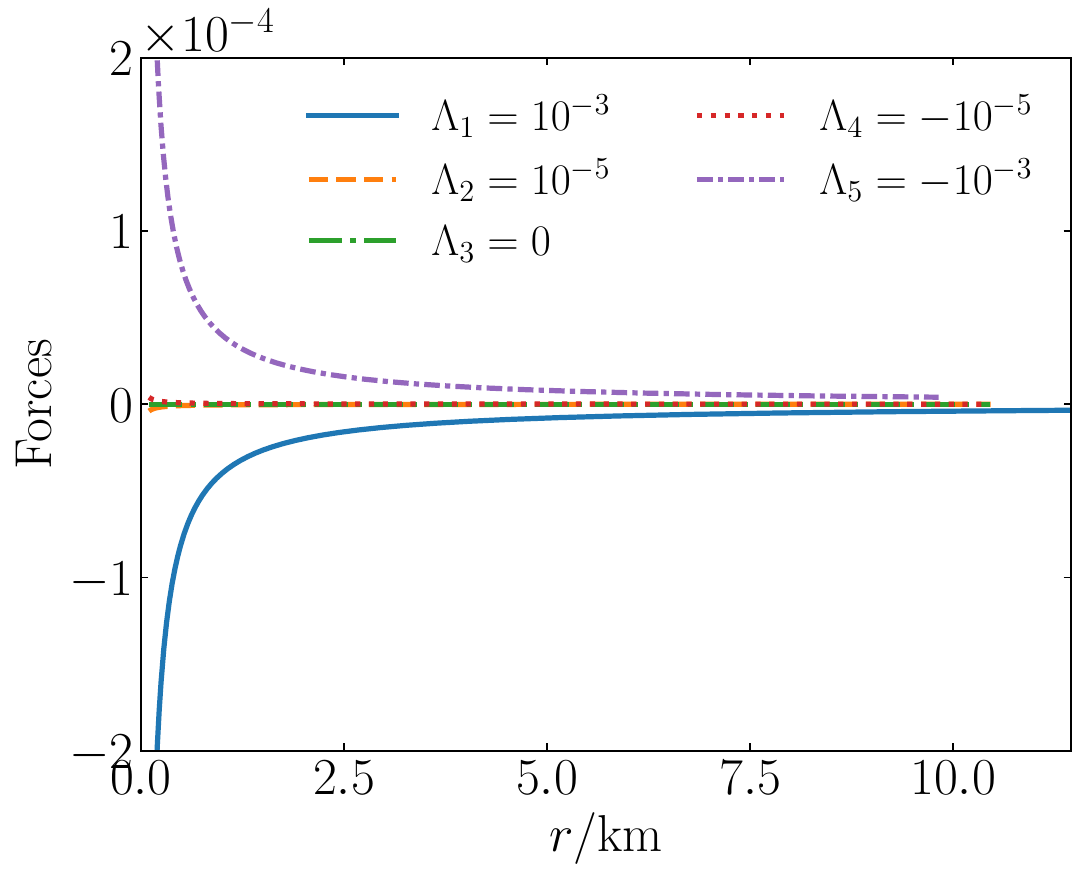}
        \caption{Total force from gravitational, hydrostatic and anisotropic effects.}
        \label{fig:forces}
    \end{subfigure}
    \hfill
    \begin{subfigure}{0.47\textwidth}
        \centering
        \includegraphics[width=\linewidth]{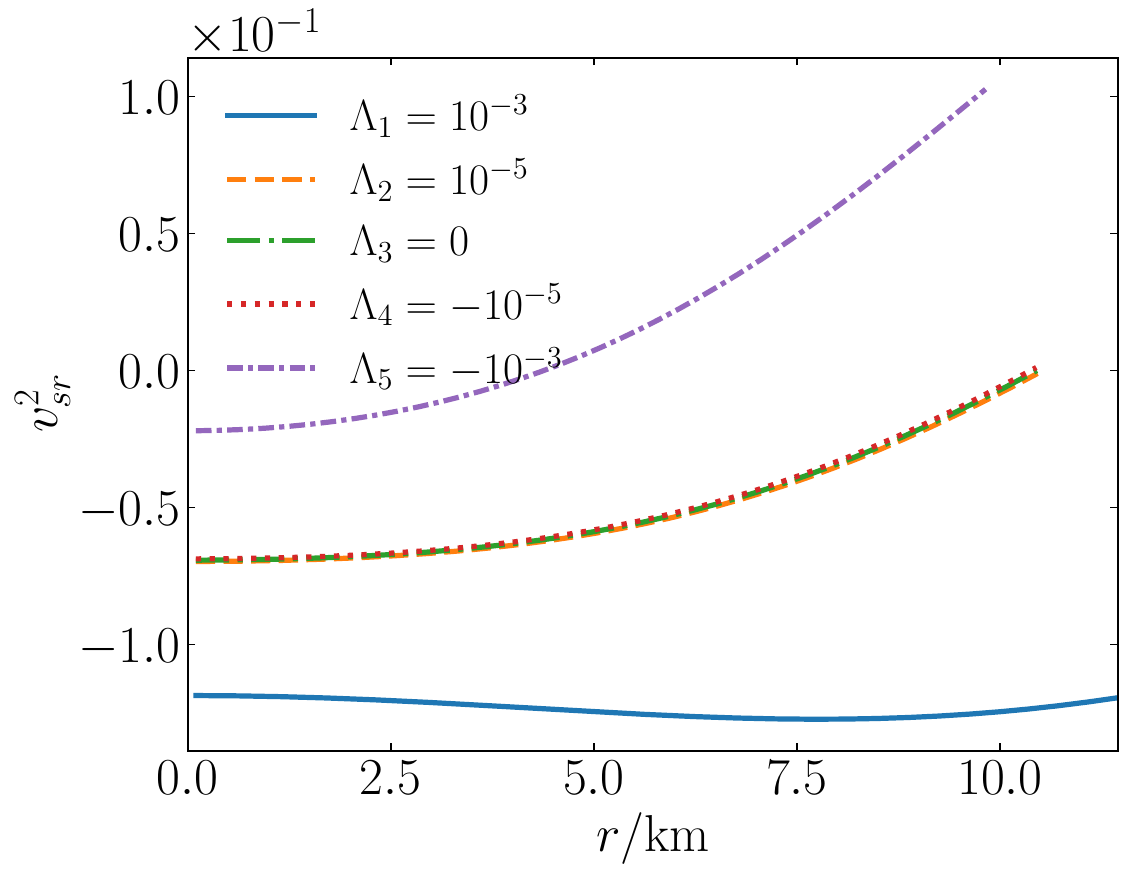}
        \caption{Radial sound speed squared, $v_{sr}^2(r)$.}
        \label{fig:vsr2}
    \end{subfigure}

    \vspace{0.2cm}

    \begin{subfigure}{0.47\textwidth}
        \centering
        \includegraphics[width=\linewidth]{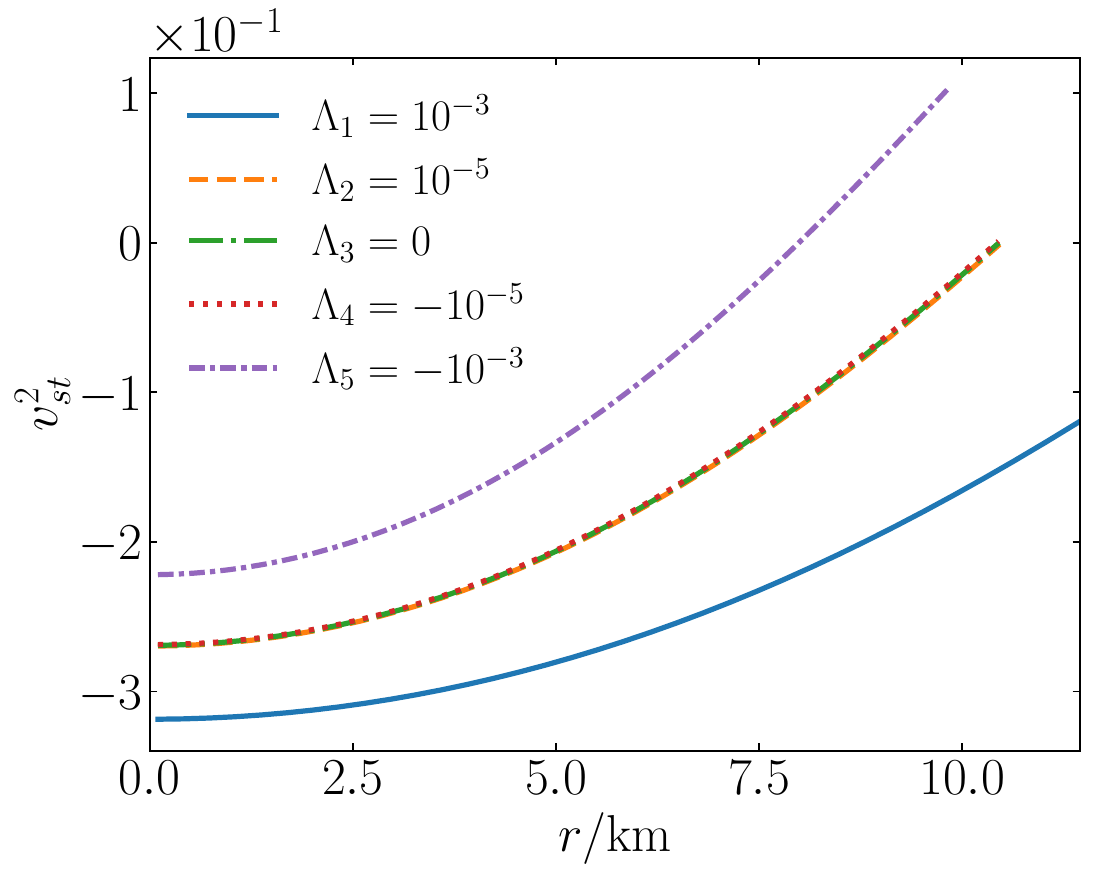}
        \caption{Tangential sound speed squared, $v_{st}^2(r)$.}
        \label{fig:vst2}
    \end{subfigure}
    \hfill
    \begin{subfigure}{0.47\textwidth}
        \centering
        \includegraphics[width=\linewidth]{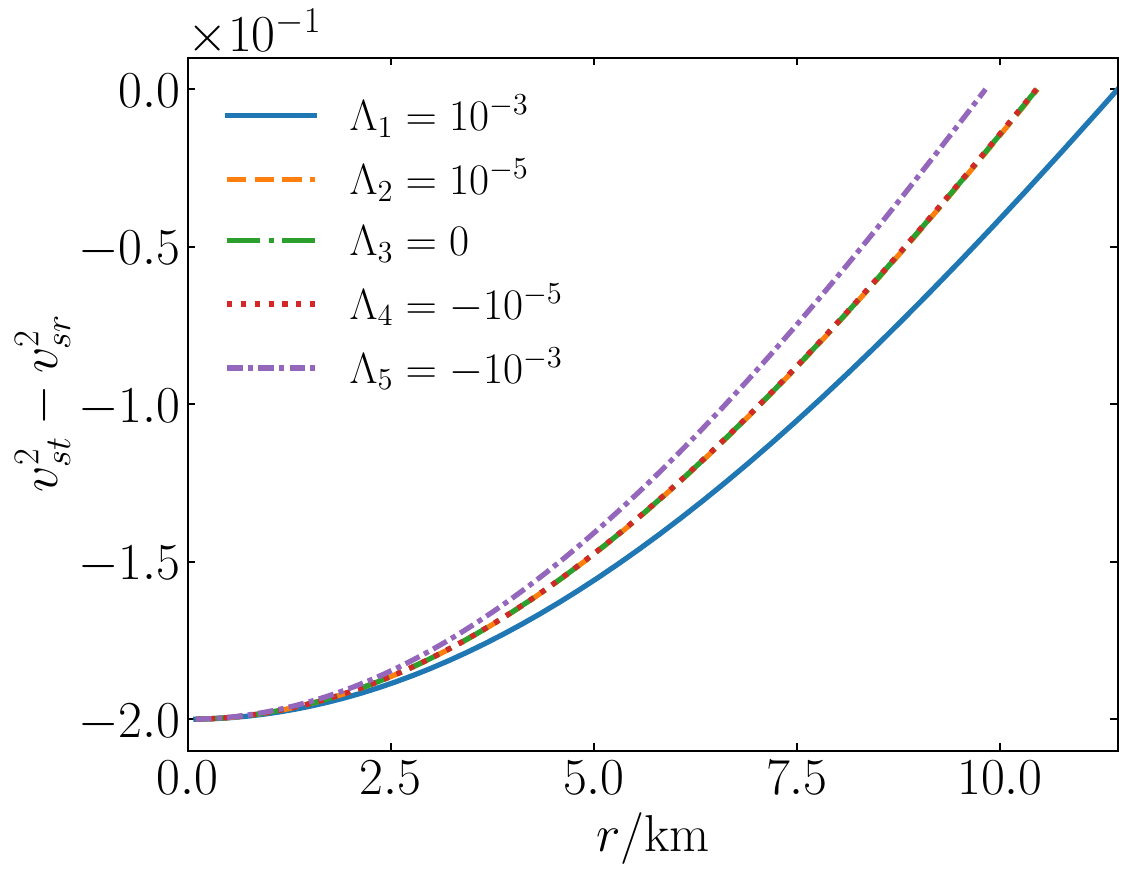}
        \caption{Difference between tangential and radial sound speeds, $v_{st}^2-v_{sr}^2$.}
        \label{fig:causality}
    \end{subfigure}

    \vspace{0.2cm}

    \begin{subfigure}{0.47\textwidth}
        \centering
        \includegraphics[width=\linewidth]{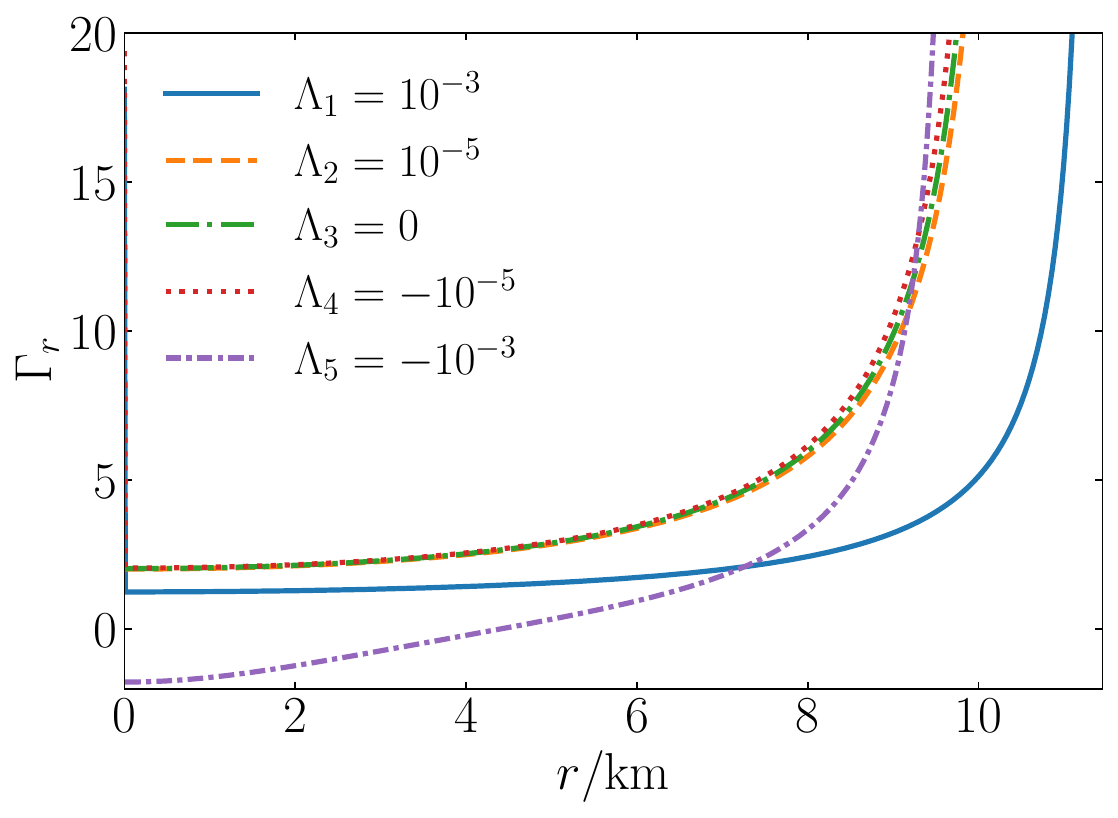}
        \caption{Radial adiabatic index $\Gamma_r(r)$.}
        \label{fig:index_r}
    \end{subfigure}
    \hfill
    \begin{subfigure}{0.47\textwidth}
        \centering
        \includegraphics[width=\linewidth]{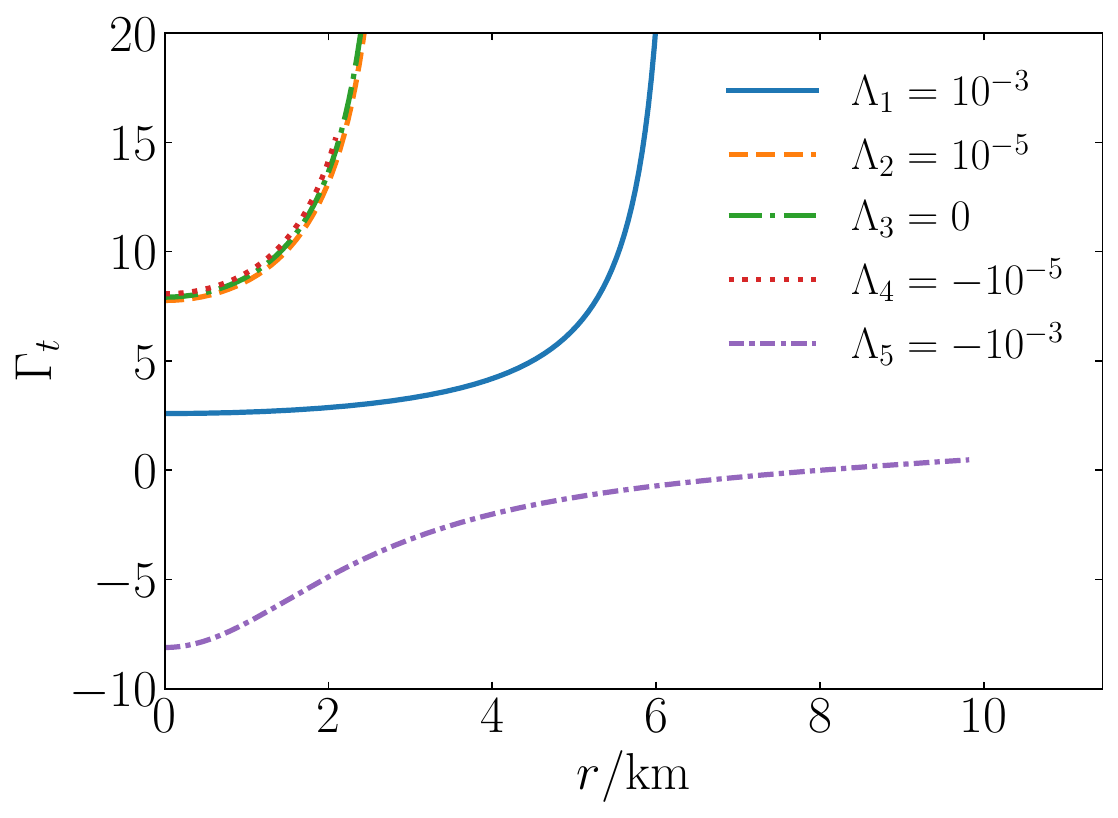}
        \caption{Tangential adiabatic index $\Gamma_t(r)$.}
        \label{fig:index_t}
    \end{subfigure}

    \caption{Equilibrium, causality and stability analysis for different values of the cosmological constant $\Lambda = 10^{-3}, 10^{-5}, 0, -10^{-5}$ and $-10^{-3}\,\mathrm{km}^{-2}$.}
    \label{fig:combined_causal_stability_forces}
\end{figure}

The hydrostatic equilibrium of the dark energy star is analyzed using the modified Tolman--Oppenheimer--Volkoff (TOV) equation, which describes the balance between the internal forces acting inside the stellar configuration. It is expressed as
\begin{equation}
    F_g + F_h + F_a = 0,
\end{equation}
where $F_g$, $F_h$ and $F_a$ denote the gravitational, hydrostatic and anisotropic forces, respectively. The radial behavior of these forces is illustrated in \autoref{fig:forces}. For the cases $\Lambda = 0$ and $\Lambda = \pm10^{-5}$, the forces remain well balanced throughout the stellar interior, indicating that the configuration satisfies hydrostatic equilibrium. However, for larger values of $\Lambda = \pm10^{-3}$, noticeable deviations from exact force balance appear, particularly toward the central regions of the star. Specifically, for $\Lambda = -10^{-3}$, the anisotropic force satisfies $F_a > 0$, whereas for $\Lambda = 10^{-3}$, $F_a < 0$. These results suggest that sufficiently large positive or negative values of the cosmological constant significantly modify the anisotropic force distribution, causing the equilibrium configuration to become increasingly sensitive to variations in \(\Lambda\).

\subsection{Causality and Dynamical Stability}

In this section, we investigate the physical viability and dynamical stability of the present dark energy star model through the causality condition, Herrera’s cracking criterion and the relativistic adiabatic index. These analyses serve as important consistency tests to determine whether the obtained solutions remain physically acceptable and stable throughout the stellar interior.

\subsubsection{Causality and Cracking}

For a physically acceptable compact star model, the radial and tangential sound speeds must satisfy the causality conditions
\[
0 \leq v_{sr}^2 \leq 1,
\qquad
0 \leq v_{st}^2 \leq 1,
\]
which ensure that the propagation of sound remains subluminal throughout the stellar interior~\cite{1992PhLA..165..206H,2007CQGra..24.4631A}. The radial and tangential sound speeds are respectively defined as
\begin{align}
    v_{sr}^2 &= \frac{d(p_r)_{eff}}{d\rho_{eff}},\\
    v_{st}^2 &= \frac{d(p_t)_{eff}}{d\rho_{eff}}.
\end{align}

The profiles of \(v_{sr}^2\) and \(v_{st}^2\) are presented in \autoref{fig:vsr2} and \autoref{fig:vst2}. The sound-speed profiles indicate partial violations of the standard causality bounds in certain interior regions, particularly for large positive values of the cosmological constant. In particular, the configuration corresponding to \(\Lambda = 10^{-3}\) exhibits the strongest deviation from the causal interval throughout the stellar interior. Meanwhile, the cases \(\Lambda = 10^{-5}\), \(\Lambda = 0\) and \(\Lambda = -10^{-5}\) violate the causality condition mainly within the inner region before approaching the allowed range near the stellar surface. The configuration with \(\Lambda = -10^{-3}\) shows comparatively improved behavior, with the violation confined primarily to the central region. These results suggest that the present phenomenological equation of state may require further refinement in order to obtain fully causal stellar configurations.

To further analyze the stability of the anisotropic matter distribution, Herrera’s cracking criterion is considered through the quantity \(v_{st}^2-v_{sr}^2\). According to this criterion, potentially stable regions satisfy the condition \( -1 \leq v_{st}^2 - v_{sr}^2 \leq 0 \). The behavior of this difference is illustrated in \autoref{fig:causality}. The profiles of \(v_{st}^2-v_{sr}^2\) remain negative and bounded well within the Herrera stability interval \((-1,0)\) throughout the stellar interior for all considered values of the cosmological constant. The quantity gradually approaches zero only near the outer boundary, supporting the absence of cracking instability despite the partial causality violations.

\subsubsection{Adiabatic Index}

The stability analysis of compact stellar configurations plays an essential role in determining the physical viability of the proposed model. In relativistic astrophysics, the adiabatic index $\Gamma$ serves as an important diagnostic quantity for examining the response of the stellar matter against infinitesimal radial perturbations. In particular, the relativistic adiabatic index measures the stiffness of the equation of state and provides information regarding the dynamical stability of anisotropic compact objects. For a stable relativistic configuration, the adiabatic index must satisfy the Chandrasekhar stability criterion \cite{1964ApJ...140..417C, 1964ApJ...140.1342C}
\[
\Gamma > \frac{4}{3},
\]
throughout the stellar interior. Therefore, to further examine the stability behavior of the present dark energy star model in the presence of positive and negative cosmological constants, we analyze the radial and tangential adiabatic indices throughout the stellar interior. The relativistic adiabatic indices corresponding to the radial and tangential directions are defined respectively as
\begin{equation}
\Gamma_r=\left(\frac{\rho_{eff}+(p_r)_{eff}}{(p_r)_{eff}}\right)\frac{d(p_r)_{eff}}{d\rho_{eff}},
\qquad
\Gamma_t=\left(\frac{\rho_{eff}+(p_t)_{eff}}{(p_t)_{eff}}\right)\frac{d(p_t)_{eff}}{d\rho_{eff}}.
\end{equation}

The profiles of the radial and tangential adiabatic indices are presented in \autoref{fig:index_r} and \autoref{fig:index_t}. The profiles for $\Lambda=0$, $\pm10^{-5}$ and $10^{-3}$ exhibit very similar behavior, suggesting that small positive or negative cosmological constants produce only minor effects on the stiffness of the stellar matter. However, for the larger negative cosmological constant case $\Lambda=-10^{-3}$, the behavior changes considerably. The radial adiabatic index $\Gamma_r$ becomes negative near the stellar center and exceeds the stability limit only beyond $r\approx7\,\mathrm{km}$, while the tangential adiabatic index $\Gamma_t$ remains below the critical value $4/3$ throughout the interior region. This indicates that larger values of the negative cosmological constant reduce the stiffness of the matter distribution and may lead to instability in the central core of the compact object.

\section{Discussion and Conclusion\label{sec:6}}

In this work, the effects of the cosmological constant on dark energy stars within the Finch--Skea spacetime framework have been analyzed. The stellar configuration was modeled as a two-fluid system composed of ordinary matter and dark energy. By employing the complexity factor formalism, an exact analytical expression for the temporal metric potential was obtained and smoothly matched to the exterior Schwarzschild--(anti--)de Sitter geometry through the appropriate boundary conditions. The resulting model allowed a detailed analysis of the influence of both positive and negative cosmological constants on the physical properties and stability behavior of compact dark energy star configurations, as summarized in \autoref{tab:summary_lambda_effects}.

\begin{table}[!ht]
\centering
\caption{
Summary of the effects of the cosmological constant on the physical properties, energy conditions and stability behavior of the present dark energy star model.
}
\scriptsize
\begin{tabular}{p{2.5cm}p{4cm}p{4cm}p{4cm}}
\toprule
\textbf{Physical quantity / criterion} 
& \textbf{Positive $\Lambda$ (SdS)} 
& \textbf{$\Lambda = 0$ (Schwarzschild)} 
& \textbf{Negative $\Lambda$ (SAdS)} \\
\midrule

Stellar radius $R$ 
& Increases 
& Reference case 
& Decreases \\

Compactness $U$ 
& Decreases 
& Moderate 
& Increases \\

Surface redshift $z_s$ 
& Decreases 
& Moderate 
& Increases \\

Central density $\rho_c$ 
& Decreases 
& Intermediate 
& Increases \\

Central anisotropy $\Delta(0)$ 
& Negative 
& Zero 
& Positive \\



Energy conditions 
(WEC, NEC, DEC, SEC)
& All satisfied  
& All satisfied 
& All satisfied\\

Hydrostatic equilibrium 
& Better balanced for small positive $\Lambda$, with deviations for large positive $\Lambda$ ($F_a < 0$)
& Well balanced 
& Better balanced for small negative $\Lambda$, with deviations for large negative $\Lambda$ ($F_a > 0$) \\

Causality condition $(0\leq v_s^2 \leq 1)$ 
& Significant deviation from the causal bounds for $\Lambda=10^{-3}$
& Partial violation within the inner region
& Improved causal behavior; deviations mainly confined to the central region for $\Lambda=-10^{-3}$\\

Herrera cracking condition 
$\left(-1 \leq v_{st}^2-v_{sr}^2 \leq 0\right)$
& Satisfied
& Satisfied 
& Satisfied  \\

Adiabatic index stability $(\Gamma > 4/3)$ 
& Stable 
& Stable 
& Stability reduced for large negative $\Lambda$; $\Gamma_t<4/3$ and $\Gamma_r<0$ near the core for $\Lambda=-10^{-3}$ \\

Overall stability 
& Physically less viable for large positive $\Lambda$ 
& Most physically acceptable configuration 
& Small negative $\Lambda$ acceptable, while large negative $\Lambda$ weakens stability \\

Overall physical effect 
& Produces less compact and less dense stars 
& Balanced compact configuration 
& Produces denser and more compact stars \\

\bottomrule
\end{tabular}
\label{tab:summary_lambda_effects}
\end{table}
The present analysis demonstrates that variations in the cosmological constant significantly influence the physical properties and stability behavior of dark energy stars within Finch--Skea spacetime. As summarized in \autoref{tab:summary_lambda_effects}, positive and negative values of $\Lambda$ produce qualitatively distinct effects on the stellar configuration. In particular, positive cosmological constants lead to larger and less compact stars with lower surface redshift, whereas negative values of $\Lambda$ produce smaller and more compact configurations accompanied by stronger gravitational effects.

Ref.~\cite{ASTEFANESEI2003594} reported that the presence of a negative cosmological constant leads to a decrease in the stellar mass. In the present analysis, the stellar mass is held constant, but the stellar radius decreases for negative values of the cosmological constant. This suggests that the cosmological constant affects not only the mass of the compact object but also its radius. Such behavior agrees with the interpretation given in Refs. \cite{ASTEFANESEI2003594,2013PhRvD..87d4003H,Cruz_2005,2021EPJC...81..610V}, where a negative cosmological constant effectively contributes to inward gravitational attraction to the stellar system. As a consequence, the stellar matter becomes more compressed, producing a smaller stellar radius. In contrast, a positive cosmological constant contributes an outward repulsive effect that tends to enlarge the stellar configuration and this influence becomes increasingly significant for larger values of positive $\Lambda$. Nevertheless, as illustrated in \autoref{fig:forces}, noticeable deviations from hydrostatic equilibrium appear when the values of the cosmological constant $|\Lambda|$ becomes too large, suggesting that the stellar configuration may become less stable in this regime. These results indicate that variations of $\Lambda$ in the exterior spacetime can influence the interior structure of the stellar configuration. 

A comparative analysis between the present study and previous investigations of dark energy stars based on Finch--Skea metric potentials for the compact star Vela X-1 is presented in \autoref{tab:summary_comparison}. From \autoref{tab:summary_comparison}, the macroscopic observables such as \(M\), \(R\), \(U\) and \(z_s\) are found to be influenced by the coupling parameter \((\alpha)\) \cite{2024ApSS.369...76D}, the model parameters \((A, B \text{ or } C)\) \cite{2024ChJPh..87..608R}, the equation of state parameter \((\omega)\) \cite{2020MPLA...3550071B} and the cosmological constant \((\Lambda)\). Therefore, it is difficult to determine conclusively which parameter contributes most significantly to the observed effects, since each study was performed using different forms of the Finch--Skea metric potential (see \autoref{tab:summary-finch-skea-metric}). This issue could be investigated further in future work to determine more definitively which parameters most strongly affect the stellar configuration. Nevertheless, the findings of the present study are expected to remain qualitatively significant even when examined using other metric potentials, which may also be explored in future investigations.

\begin{table}[!ht]
\centering
\caption{
Comparison of macroscopic observables for Finch--Skea dark energy star models under different values of the cosmological constant \(\Lambda\), coupling parameter \(\alpha\), equation of state parameter \(\omega\) and model parameter \(C\).
}

\scriptsize
\setlength{\tabcolsep}{3pt}
\renewcommand{\arraystretch}{0.92}

\resizebox{\textwidth}{!}{%
\begin{tabular}{lcccccccccc}
\toprule
Exterior Spacetime 
& $\Lambda$ (km$^{-2}$)
& $\alpha$
& $A$ (km$^{-2}$)
& $B$
& $C$ (km$^{-2}$)
& $\omega$
& $M\,(M_{\odot})$
& $R$ (km)
& $U$ 
& $z_s$
\\
\midrule

\multicolumn{11}{c}{\textbf{Effect of cosmological constant $\Lambda$}~[This study]}\\
\midrule

SdS
& $10^{-3}$
& $\alpha=1.0$
& 0.00140
& 0.53310
& $-$
& $-1$
& 1.77
& 11.45
& 0.228
& 0.356
\\

SdS
& $10^{-5}$
& $\alpha=1.0$
& 0.00227
& 0.47192
& $-$
& $-1$
& 1.77
& 10.46
& 0.250
& 0.414
\\

Schwarzschild
& $0$
& $\alpha=1.0$
& 0.00228
& 0.47140
& $-$
& $-1$
& 1.77
& 10.45
& 0.250
& 0.414
\\

SAdS
& $-10^{-5}$
& $\alpha=1.0$
& 0.00229
& 0.47088
& $-$
& $-1$
& 1.77
& 10.44
& 0.250
& 0.415
\\

SAdS
& $-10^{-3}$
& $\alpha=1.0$
& 0.00309
& 0.42592
& $-$
& $-1$
& 1.77
& 9.82
& 0.266
& 0.462
\\

\midrule

\multicolumn{11}{c}{\textbf{Effect of coupling parameter $\alpha$}~\cite{2024ApSS.369...76D}}\\
\midrule

SdS
& $1.8 \times 10^{-4}$
& $\alpha=0.0$
& 0.452126
& 0.0391468
& 0.0131686
& $-1$
& 2.58
& 9.56
& 0.270
& 0.476
\\

SdS
& $1.8 \times 10^{-4}$
& $\alpha=0.1$
& 0.369145
& 0.0851785
& 0.0131686
& $-1$
& 2.35
& 9.56
& 0.246
& 0.402
\\

SdS
& $1.8 \times 10^{-4}$
& $\alpha=0.2$
& 0.299993
& 0.1235380
& 0.0131686
& $-1$
& 2.15
& 9.56
& 0.225
& 0.349
\\

SdS
& $1.8 \times 10^{-4}$
& $\alpha=0.3$
& 0.241480
& 0.155997
& 0.0131686
& $-1$
& 1.99
& 9.56
& 0.208
& 0.308
\\

SdS
& $1.8 \times 10^{-4}$
& $\alpha=0.4$
& 0.191326
& 0.183818
& 0.0131686
& $-1$
& 1.85
& 9.56
& 0.193
& 0.276
\\

\midrule

\multicolumn{11}{c}{\textbf{Effect of model parameter $C$ (and $\alpha$)}~\cite{2024ChJPh..87..608R}}\\
\midrule

Schwarzschild
& $0$
& $\alpha = 0.40 - 1.20$
& $0.00055 - (-0.00024)$
& $0.67207 - 0.79480$
& 0.005
& $-1$
& 1.77
& 12.20
& 0.214
& 0.304
\\

Schwarzschild
& $0$
& $\alpha = 0.40 - 1.20$
& $0.00106 - (-0.00020)$
& $0.59669 - 0.71914$
& 0.010
& $-1$
& 1.77
& 10.20
& 0.256
& 0.428
\\

Schwarzschild
& $0$
& $\alpha = 0.40 - 1.20$
& $0.00152 - (-0.00007)$
& $0.54325 - 0.66598$
& 0.015
& $-1$
& 1.77
& 9.28
& 0.281
& 0.513
\\

Schwarzschild
& $0$
& $\alpha = 0.40 - 1.20$
& $0.00194 - 0.00008$
& $0.50264 - 0.62588$
& 0.020
& $-1$
& 1.77
& 8.69
& 0.301
& 0.587
\\

\midrule

\multicolumn{11}{c}{\textbf{Effect of equation of state parameter $\omega$}~\cite{2020MPLA...3550071B}}\\
\midrule

Schwarzschild
& $0$
& $-$
& $-$
& $-$
& $-$
& $-0.37$
& 1.77
& 10.736
& 0.273
& 0.357
\\

Schwarzschild
& $0$
& $-$
& $-$
& $-$
& $-$
& $-0.41$
& 1.77
& 10.652
& 0.2645
& 0.348
\\

Schwarzschild
& $0$
& $-$
& $-$
& $-$
& $-$
& $-0.47$
& 1.77
& 10.611
& 0.2743
& 0.337
\\

\bottomrule
\end{tabular}
}

\label{tab:summary_comparison}
\end{table}

Apart from the study of Ref.~\cite{2024ApSS.369...76D}, in which the stellar radius was kept constant throughout the model, the zero cosmological constant model of Ref.~\cite{2024ChJPh..87..608R} appears to be the least constrained, since its model parameters can be adjusted more freely to obtain a stellar radius within the predicted range of \(9.56 \pm 0.08\) km \cite{10.1093/mnras/stt401} for Vela X-1. In contrast, the model of Ref. \cite{2020MPLA...3550071B} is not expected to approach the predicted radius significantly, since the stellar radius decreases by approximately $0.125\,\mathrm{km}$ for every $0.1$ decrease in the equation of state parameter $\omega$ within the quintessence regime, unless the parameter enters the phantom regime. In the present model, the stellar radius is constrained by the requirement of hydrostatic equilibrium, since large values of $|\Lambda|$ lead to deviations from equilibrium, thereby preventing the radius from reaching the predicted value. This limitation may be improved if the metric potential $g_{rr}$ is assumed in the form proposed in Refs. \cite{2024ApSS.369...76D,2024ChJPh..87..608R}, where the stellar radius depends not only on the cosmological constant but also on additional model parameters, providing greater flexibility in fitting the observed stellar properties.

The present analysis is based on a phenomenological dark energy equation of state and a prescribed interior geometry. Consequently, additional investigations involving alternative equations of state, perturbative analyses and observational constraints would be valuable for establishing the astrophysical viability of the model more rigorously. Future studies may investigate whether the exterior properties of the spacetime, such as null and timelike geodesics, are influenced by variations in the parameters governing the interior structure of the stellar configuration. For example, Ref.~\cite{2017EPJC...77..123C} showed that the effective potential increases for negative values of the cosmological constant and decreases for positive values relative to the case \(\Lambda = 0\). However, the manner in which the interior properties affect the exterior behavior was not explicitly examined. A systematic investigation of this relationship would therefore be of considerable interest and may help clarify the coupling between the interior and exterior spacetime properties. The present study therefore suggests that the cosmological constant may play a nontrivial role not only in cosmology but also in determining the equilibrium structure and observable properties of compact self-gravitating systems. These findings motivate further investigations of dark energy stars in more general interior geometries and alternative gravitational frameworks.


\section*{Declaration of competing interest}
The authors declare that they have no known competing financial interests or personal relationships that could have appeared to influence the work reported in this paper.

\section*{Acknowledgements}
M.A.A. acknowledges the use of the following packages in this work: {\fontfamily{pcr}\selectfont SYMPY} \cite{10.7717/peerj-cs.103} for systematic symbolic derivations of the relevant parameters, together with {\fontfamily{pcr}\selectfont NUMPY} \cite{2011CSE....13b..22V} and {\fontfamily{pcr}\selectfont MATPLOTLIB} \cite{2007CSE.....9...90H} for graphical plotting.

\section*{Declaration of generative AI and AI-assisted technologies in the manuscript preparation process}
During the preparation of this work, the author(s) used OpenAI’s ChatGPT-5.5 to assist with substantial rephrasing, sentence restructuring and content enhancement in certain sections. After using this tool, the author(s) reviewed and edited the content as needed and take full responsibility for the content of the published article.

\bibliographystyle{elsarticle-num-names}
\bibliography{References}

\appendix
\renewcommand{\thetable}{A\arabic{table}}
\setcounter{table}{0}

\begin{sidewaystable}[p]
\centering
\caption{
Summary of metric potentials $e^{\lambda(r)}$ and $e^{\nu(r)}$ employed in various Finch--Skea-based compact star models across different gravitational frameworks. The constants $A$, $B$, $C$, $D$, $\mathfrak{N}$, $a$, $c$, $n$, $\gamma$ and $\nu_0$ are defined according to the respective references.
}\label{tab:summary-finch-skea-metric}%
\scriptsize\begin{tabular}{lllll}
\hline
Ref. & $e^{\lambda(r)}$ & $e^{\nu(r)}$ & Theory of gravity & Compact star\\
\hline
\cite{2020MPLA...3550071B} & $1 + \dfrac{r^2}{R_*^{2}}$ & $-$ & Einstein & Dark energy stars\\
\cite{2024ApSS.369...76D} & $1 + Cr^2$ & $(B - A\sqrt{1 + Cr^2})\cos{(\sqrt{1 + Cr^2}) + (A + B\sqrt{1 + Cr^2})\sin{(\sqrt{1 + Cr^2})}}$ & Einstein (+ cosmological constant) & Dark energy stars\\
\cite{2024ChJPh..87..608R} & $1 + Cr^2$ & $\left(B + \dfrac{A(1 + Cr^2)^{3/2}}{3C}\right)^2$ & Einstein & Dark energy stars\\
\cite{2022arXiv220613943M} & $1 + acr^2$ & $A^2c^2(1+ac^2r^4)^{2A}(1+acr^2)e^{\frac{(\omega + 1)\sqrt{a}\arctan{cr^2\sqrt{a}}}{2}}$ & Einstein–Maxwell & Dark energy stars\\
This Study & $1 + \dfrac{r^2}{R_*^{2}}$ & $\left(B + \dfrac{A \left(R_*^{2} + r^{2}\right)^{3/2}}{3 R_*}\right)^{2}$ & Einstein (+ cosmological constant) & Dark energy stars\\
\cite{2025ApSS.370...46I} 
& $1 + Cr^2$ 
& $\left(A + \dfrac{B\sqrt{C}\,r^2}{2}\right)^2$ 
& $f(\mathcal{R}, \mathcal{T})$ (+ cosmological constant)
& Anisotropic stars \\
\cite{2023ChPhC..47a5104S} & $1 + Cr^2$ & $\left(A + \dfrac{Br^2\sqrt{C}}{2}\right)^2$ & Einstein & Anisotropic stars\\
\cite{2023GrCo...29..206J} & $1 + \dfrac{r^2}{R_*^{2}}$ & $\left(C\left(1 + \dfrac{r^2}{R_*^{2}}\right)^{5/4} + D\left(1 + \dfrac{r^2}{R_*^{2}}\right)^{1/4}\right)^2$ & Einstein & Anisotropic stars\\
\cite{2020CQGra..37g5017D} & $\sqrt{1 + Cr^2}$ & $(B - A\sqrt{1 + Cr^2})\cos{(\sqrt{1 + Cr^2}) + (A + B\sqrt{1 + Cr^2})\sin{(\sqrt{1 + Cr^2})}}$ & Einstein–Maxwell in higher dimensions & Anisotropic stars\\
\cite{2021JApA...42...74B} & $\left(1 + \dfrac{r^2}{R_*^{2}}\right)^2$ & $\left(D + \dfrac{C(r^2 + R_*^{2})^{1+\frac{n}{2}}}{2 + n}\right)^2$ & Einstein & Anisotropic stars\\
\cite{2023ChJPh..81..362D} & $1 + \dfrac{r^2}{R_*^{2}}$ &  $\left(C\sqrt{r^2 + R_*^{2}} + D(r^2 + R_*^{2})\right)^2$ & Einstein & Anisotropic stars\\
\cite{2013GReGr..45..717B} & $1 + \dfrac{r^2}{R_*^{2}}$ & $\left(D - A\sqrt{1 + \dfrac{r^2}{R_*^{2}}}\right)\cos{\sqrt{1 + \dfrac{r^2}{R_*^{2}}} + \left(D\sqrt{1 + \dfrac{r^2}{R_*^{2}}} + A\right)\sin{\sqrt{1 + \dfrac{r^2}{R_*^{2}}}}}$ & Einstein in (2 + 1) dimension & Perfect fluid stars\\
\cite{2021IJGMM..1850160B} & $1 + ar^2$ & $((B - A\sqrt{1 + ar^2})\cos{(\sqrt{1 + ar^2}) + (A + B\sqrt{1 + ar^2})\sin{(\sqrt{1 + ar^2})}})^2$ & $f(\mathcal{R}, \mathcal{T})$ & Perfect fluid stars\\

\cite{MAJEED2022101802} & $1 + ar^2$ & $e^{A - 4B\rho_0 + c_0}$ where $A = \dfrac{ar^2}{2}$ and $B = \left(\dfrac{r^2}{2} + \dfrac{ar^4}{4}\right)(2\pi + \gamma)$ & $f(\mathcal{R}, \mathcal{T})$ & Gravastars \\
\cite{2023MPLA...3850123S} & $1 + \mathfrak{N}r^2$ & $-$ & $f(\mathfrak{R},\mathcal{T}^2)$& Gravastars\\
\cite{2013IJTP...52.3319K} & $1 + \dfrac{r^2}{R_*^{2}}$ & $e^{\nu_0}(R_*^{2}+r^2)^{\frac{1}{3}}e^{-\frac{8\pi Br^2}{3R_*^{2}}\left(2R_*^{2} + r^2 - \frac{1}{4\pi B}\right)}$ & Einstein & Quark stars\\
\hline
\end{tabular}
\end{sidewaystable}






\end{document}